\documentclass[]{aa}
\usepackage[varg]{txfonts}
\usepackage{natbib}
\usepackage{xcolor}
\usepackage{libertine}
\usepackage{siunitx}
\usepackage[normal]{subfigure}
\usepackage{enumitem}
\bibpunct{(}{)}{;}{a}{}{,} % to follow the A&A style
\usepackage{comment}
\usepackage{graphicx}

\newcommand{\ngal}{{N_{\rm gal}}}
\newcommand{\Sx}{{S_{\rm X}}}
\newcommand{\Sxprime}{{S'_{\rm X}}}
\newcommand{\Sstd}{{S_{\rm std}}}
\newcommand{\Strue}{{S_{\rm true}}}

\begin{document}

\title{Biases in galaxy cluster velocity dispersion and mass estimates in the small number of galaxies regime}
\titlerunning{GC velocity dispersion and mass estimators}

\author{A.~Ferragamo\inst{1,2}
     \and J.~A.~Rubi\~{n}o-Mart\'{\i}n\inst{1,2}
     \and J. Betancort-Rijo\inst{1,2}
     \and E.~Munari\inst{3,4}
     \and B.~Sartoris\inst{3,4}
     \and R.~Barrena\inst{1,2}}

\institute{Instituto de Astrof\'{\i}sica de Canarias (IAC), C/ V\'{\i}a L\'actea s/n, E-38205 La Laguna, Tenerife, Spain
  \and Universidad de La Laguna, Departamento de Astrof\'{\i}sica, C/ Astrof\'{\i}sico Francisco S\'anchez s/n, E-38206 La Laguna, Tenerife, Spain  
  \and Dipartimento di Fisica, Sezione di Astronomia, Universit\`a di Trieste, Via Tiepolo 11, I-34143 Trieste, Italy
  \and INAF/Osservatorio Astronomico di Trieste, Via Tiepolo 11, I-34143 Trieste, Italy}

\abstract {}
{We present a study of the statistical properties of three velocity dispersion
  and mass estimators, namely biweight, gapper and standard deviation, in the
  small number of galaxies regime ($N_{\rm gal} \le 75$). }
{Using a set of 73 numerically simulated galaxy clusters, we first characterise
  the statistical bias and the variance for each one of the three estimators
  (biweight, gapper, and standard deviation), both in the determination of the
  velocity dispersion and the dynamical mass of the clusters via the
  $\sigma$--$M$ relation. These results are used to define a new set of unbiased
  estimators, that are able to correct for those statistical biases with a
  minimal increase of the associated variance. We also used the same set of
  numerical simulations to characterise two other physical biases affecting the
  estimates: the impact of velocity segregation in the selection of cluster
  members, and the impact of using cluster members within different physical
  radii from the cluster center. }
{ The standard deviation (and its unbiased counterpart) is the lowest variance
  estimator when compared to the biweight and the gapper. We find that, due to
  the effect of velocity segregation, the selection of galaxies within the
  sub-sample of the most massive galaxies in the cluster introduces a $2\,$\%
  bias in the velocity dispersion estimate when calculated using a quarter of
  the most massive cluster members. We also find a dependence of the velocity
  dispersion estimate on the aperture radius as a fraction of $R_{200}$,
  consistent with previous results in the literature.}
{The proposed set of unbiased estimators effectively provide a correction of the
  velocity dispersion and mass estimates from those statistical and physical
  effects discussed above, in the small number of cluster members regime. By
  applying these new estimators to a subset of simulated observations, we show
  that they can retrieve bias-corrected values for both the mean velocity
  dispersion and the mean mass, being the standard deviation the one with the
  lowest variance. Although for a single galaxy cluster the
    statistical and physical effects discussed here are comparable or
    slightly smaller than the bias introduced by interlopers,
    they will be of relevance when dealing with ensemble properties
    and scaling relations for large number of clusters. }


\maketitle

%%%%%%%%  SECTION 1: INTRODUCTION  %%%%%%%%
\section{Introduction}
\label{sec:introduction}

Galaxy clusters (GCs) are tracers of the evolution of structures throughout the
history of the Universe. Cosmological parameters, such as the matter density
$\Omega_{\rm m}$ and the amplitude of matter fluctuation $\sigma_8$, are very
sensitive to the abundance of GCs per unit of mass over time
\citep[e.g.][]{voit05,allen11,planck13_count,planck15_count}.

Given that it is not possible to weigh GCs directly, we need to use mass proxies
based on other mass-related observables through scaling relations
\citep[e.g.][]{stanek10, kravtsov12}. Nowadays, there are several of these
observational proxies that are used to obtain the total cluster mass: X-ray
intracluster emission, weak lensing models, and, very recently, the
Sunyaev--Zel'dovich (SZ) effect \citep{SZ70}.
In this last method, ground-based telescopes and instruments, as the Atacama
Cosmology Telescope (ACT) \citep{atacama}, the South Pole Telescope (SPT)
\citep{SPT} or the NIKA2 instrument at IRAM \citep{NIKA2}, and space missions as
the ESA's Planck satellite \citep{planck13_count,planck15_map}, are opening new
windows for to the detection of GCs through their SZ effect.

The integrated amplitude of the inverse Compton parameter along the line of
sight, $Y$, is a good proxy for retrieving the mass of hot intra-cluster gas
\citep[e.g.][]{Arnaud10, planck13_count, planck15_count, ruel14,
  sifon16}). Moreover, the mass can be estimated using the GC luminosity in the
X-ray provided by surveys performed by the XMM satellite. Finally, in the
visible range, it is possible to infer the GC mass by studying the deformation
of background galaxy shapes due to weak lensing
\citep[e.g.][]{zitrin15,umetsu15}, computing their richness
\citep[e.g.][]{popesso07,rozo09}, or by estimating the GC velocity dispersion by
measuring the radial velocity of galaxy members
\citep[e.g.][]{biviano06}. Unfortunately, each of these observables suffers from
biases that lead to inaccurate estimates of the mass. A precise characterisation
of these biases has become of special importance in recent years due to the
recent Planck results on cluster counts \citep{planck13_count, planck15_count},
showing that some cosmological parameters, especially $\sigma_8$, inferred from
X-ray observations and SZ mass estimates are in mild tension with those deduced
from the study of the primordial anisotropies of the Cosmic Microwave Background
(CMB).

Several authors have used the velocity dispersion mass proxy to study and
characterise scaling relations between dynamical and the SZ mass \citep{ruel14,
  sifon16, amodeo17}. In this kind of study, it is necessary to quantify the
velocity dispersion of a large number of clusters, so observational programs
with limited telescope time are forced to obtain radial velocities for a low
number of members for each cluster target. Several techniques have been proposed
to minimize the impact of the low number of cluster members for determining
accurate velocity dispersion. \cite{beers90} have studied the behaviour of
different locations and scale estimators in the presence of deviation from
Gaussianity and a reduced sample of galaxies. They focused their work on the
robustness and, in particular, on the efficiency of those statistical tools. In
particular the biweight \citep{tukey58} became the standard for estimating the
velocity dispersion of galaxy samples of almost all sizes because of its
robustness and high efficiency. Over the last decade, the development of
$N$-body and hydro-dynamical simulations has given us the possibility of testing
velocity dispersion estimators directly on samples that mimic observations of
GCs.

The correct choice of an appropriate scale estimator can prevent the occurrence
of strong deviation from the actual velocity dispersion, even if GCs are sampled
with a few galaxy members. Unfortunately, these poor galaxy samples often
contain only bright galaxies owing to observational limitations. There are
several studies that take into account velocity segregation of galaxies due to
their luminosity and spectral type \citep[e.g.][]{ biviano92, goto05,
  barsanti16, bayliss17}. Dynamical friction \citep{chandrasekhar43} could be
one of the causes of the underestimation of the velocity dispersion
\citep[e.g.][]{merritt85,koylan-kolchin08, wetzel10}.  We present an analysis of
three different velocity dispersion estimators: biweight \citep{tukey58}, gapper
\citep{gapper}, and standard deviation. Using 73 simulated GCs, we test their
statistical properties when they are applied to samples made up of few galaxy
members or are contaminated by interlopers. We pay particular attention to the
case of samples containing only massive GC members.

This work is organized as follows. In Section \ref{sec:sim}, we give a brief
description of the simulations used in this paper. In Section \ref{sec:recipe}
we present the recipe to an unbiased estimates of GC velocity dispersion and
mass. In Section \ref{sec:bias}, we present a comparison of the bias and
variance for three scale estimators---biweight, gapper, and standard
deviation---as a function of the number of galaxy members considered. In Section
\ref{subsec:bias_inter}, we test the robustness of the three estimators in the
case galaxy samples containing interlopers. In Section \ref{sec:bias2}, we
quantify the effect induced on the velocity dispersion estimate by sampling
galaxy members in only a fraction of visible objects and within apertures
different from $R_{200}$. In Sections \ref{sec:s_mass_bias} and
\ref{subsec:mass_bias}, we describe how the mass can be biased even in presence
of a unbiased velocity dispersion. In Section \ref{subsec:M_recipe} we apply the
correction to a set of simulated observations based on the Planck PSZ1 optical
follow-up \citep{nostro16, rafa18} and we give the recipe to correct for the
biases in velocity dispersion and mass estimates. Finally, we present our
conclusions in Section \ref{subsec:end}.

Throughout this paper, we define $R_{200}$ as the radius within which the mean
cluster density is $200$ times the critical density of the Universe at redshift
$z$. The mass $M_{200} = (4\pi/3) 200 \rho_c(z) R_{200}^3$ is the total mass
within $R_{200}$. Other quantities with the subscript $200$ have to be
considered as evaluated at, or within $R_{200}$.

%%%%%%%%%  SECTION 2: Simulations  %%%%%%%%
\section{Simulations}
\label{sec:sim}

\begin{figure*}
\begin{center}
\includegraphics[width=0.33\textwidth]{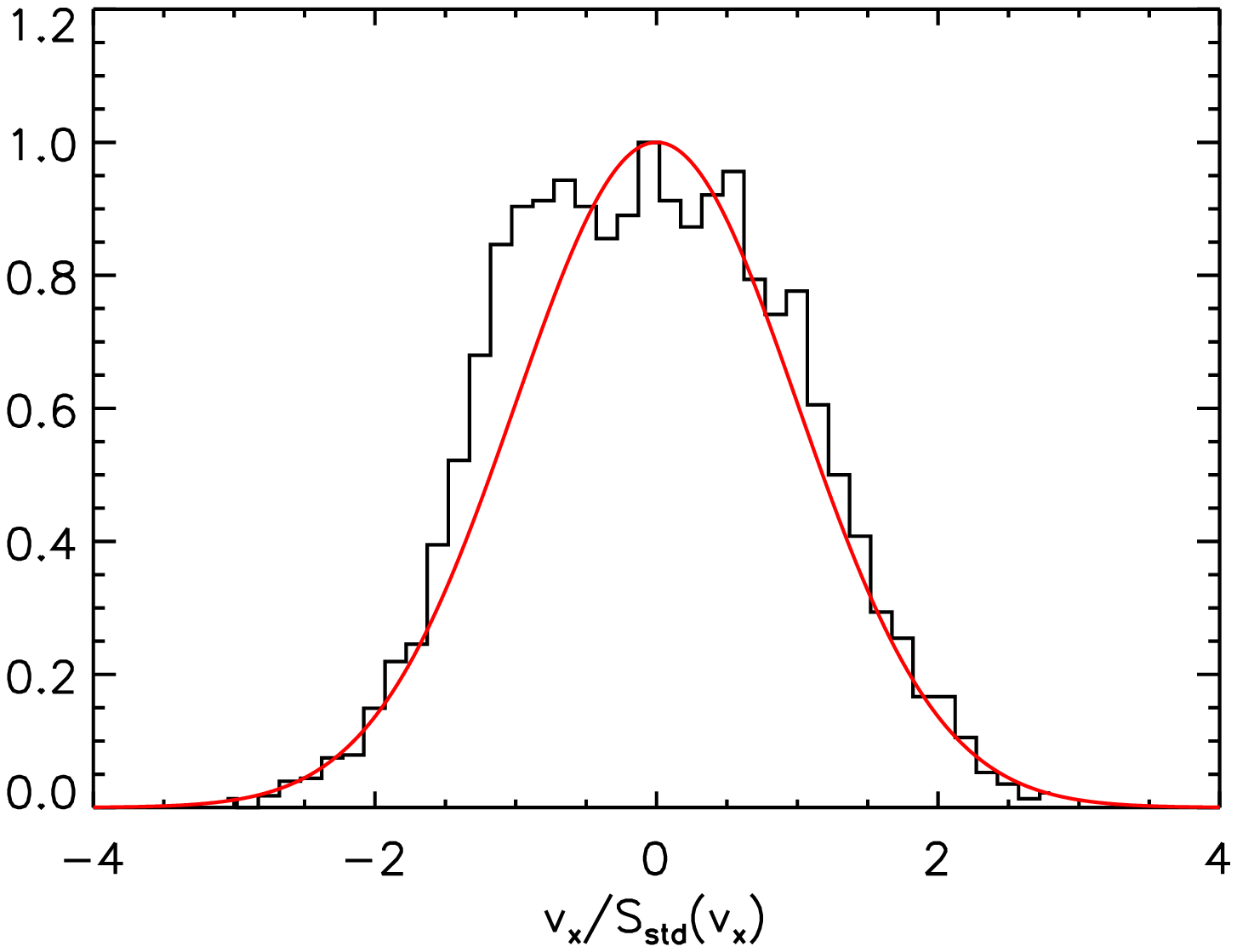}
\includegraphics[width=0.33\textwidth]{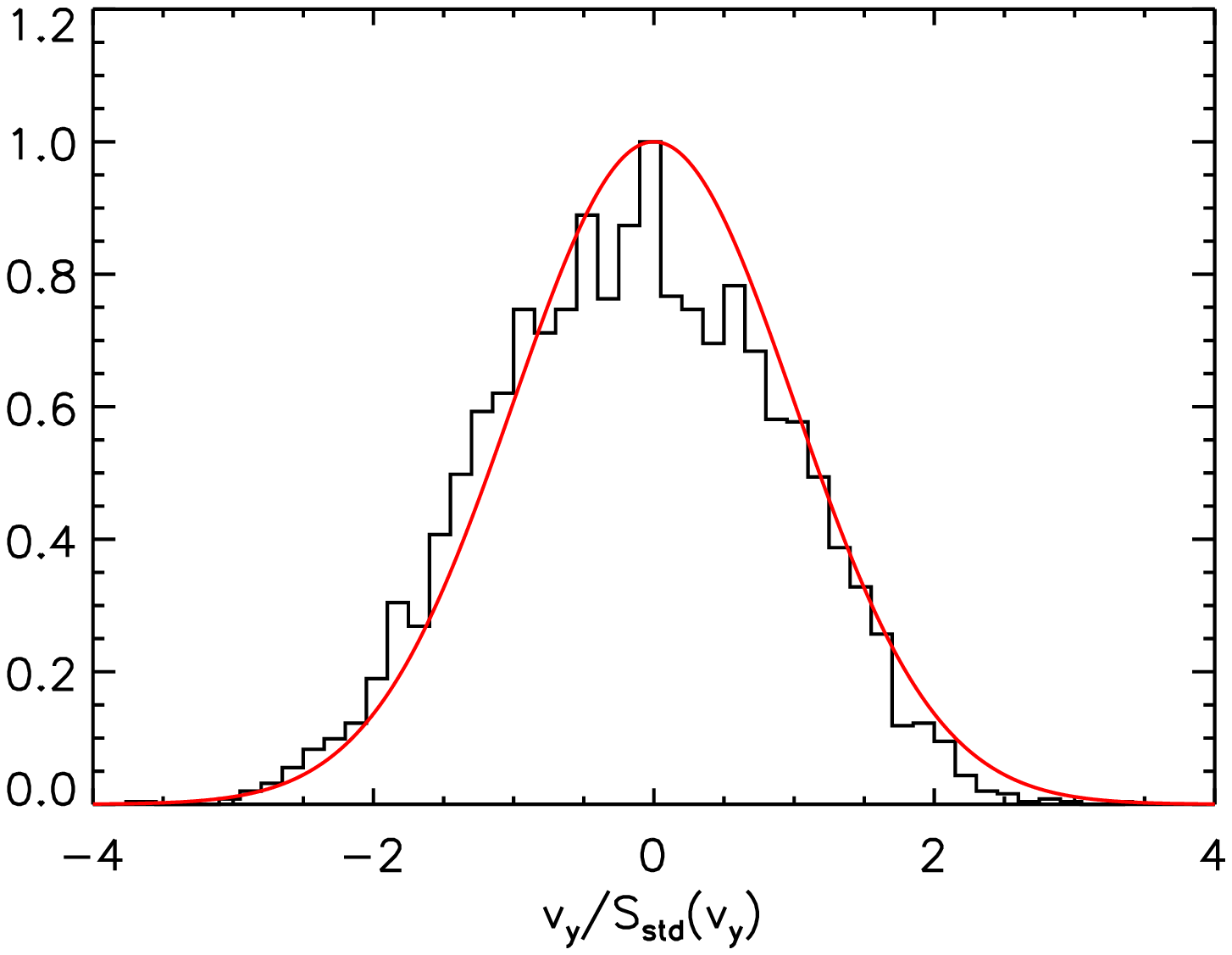}
\includegraphics[width=0.33\textwidth]{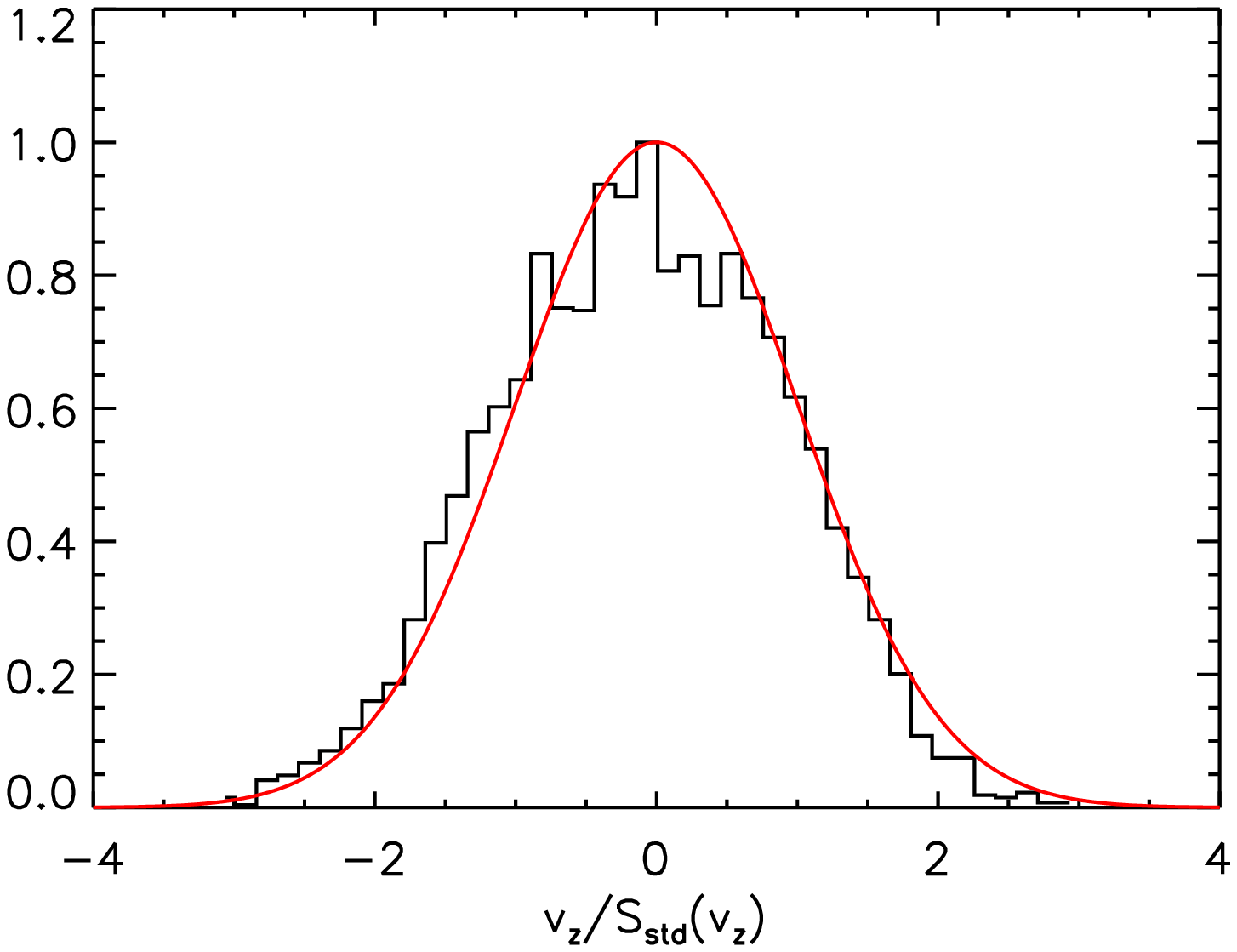}
\end{center}
\caption{Histograms showing the distribution of individual line-of-sight
  velocities of all cluster members contained in a cylinder of
    projected radius $R_{200}$ along each selected axis, for all the
  73 simulated clusters, and for three possible orientations of the simulation box
  along the line of sight (x, y and z axes). The red line corresponds to a
  Gaussian profile with the same mean and variance of the full sample. }
  \label{fig:histos}
\end{figure*}

In order to carry out the proposed analyses, we use a sample of 73 simulated
massive clusters selected from the simulations described in
\citet{munari13}. The original, full sample contains about 300 cluster-sized or
group-sized structures with masses $M_{200}>10^{13}M_\odot$. Here, our selected
sample corresponds to all clusters located at five redshifts ($z = 0.12, 0.36,
0.46, 0.6, 0.82$), and with masses $M_{200} > 2 \times 10^{14} M_\odot$.

The simulations were generated in $29$ Lagrangian regions, centred around the
massive haloes identified in a parent, large-volume simulation box of
$1$\,h$^{-1}$\,Gpc a side, and then re-simulated with higher resolution. The
simulation starts in the initial conditions described in \cite{bonafede11}, and
was carried out in two subsequent steps, at different resolutions. The entire
simulation was performed using the hydro-dynamical {\sc GADGET-3} code
\citep{springel01}. Gravitational forces are simulated using the TreeePM method,
in which the Plummer-equivalent softening length $\epsilon = 5$\,h$^{-1}$\,kpc
is assumed in physical units for $z<2$ and fixed in comoving units for
$z>2$. This simulation follows the evolution of $1024^3$ dark matter (DM)
particles with mass $m_{\rm DM}=8.47 \times 10^8$\,h$^{-1}$\,$M_\odot$ and the
same number of gas particles with initial mass $m_{\rm gas}=1.53\times
10^8$\,$M_\odot$, assuming the $\Lambda$-CDM cosmological model with
$\Omega_{\rm DM}=0.24$, $\Omega_{\rm b}=0.04$, $\Omega_\Lambda=0.72$, $H_0 =
72$\,km\,s$^{-1}$\,Mpc$^{-1}$, $\sigma_8=0.8$, and $n_{\rm S}=0.96$.

Concerning the simulation model, we use the ``AGN'' simulation set
  described in \citet{munari13}. This is a set of radiative simulations which
  account for the effect of star formation and the feedback triggered by both
  supernova explosions (SNe) and active galactic nuclei (AGN). Radiative cooling
  rates are computed following \citet{Wiersma09}. The prescription by
  \cite{Tornatore07} is used to include metal enrichment of the intra-cluster
  medium (ICM) due to SNe (both type II and Ia) and asymptotic giant branch
  (AGB) stars, taking also into account the \cite{Chabrier03} initial mass
  function (IMF) for the stars population. For a more accurate description of
  the simulations and the different prescriptions, see \cite{munari13} and
  \cite{rasia15}.

The bounded structures were identified first through a Friend-of-Friend
  (FoF) algorithm. Then, the identifications were refined using the {\sc
    SUBFIND} algorithm \citep{Springel_Yoshida01, dolag09}. The DM sub-haloes
  identified in this way that contains stellar structure are considered
  "galaxies". In analogy with \cite{munari13}, in this work we consider only
  galaxies containing a bounded stellar mass $3 \times 10^{9} \,M_{\odot}$. This
  choice guarantees that we retain all sub-haloes more massive than $\sim
  10^{11}\,M_{\odot}$. In total, there are $105,196$ galaxies in our sample of 73
  clusters, being $17,433$ within the $R_{200}$ radius. Thus, on
  average we have $1440$ galaxies per cluster, being  $239$ inside
  $R_{200}$.
%

%%%%%%%%  SECTION 3: Recipe for a  bias-corrected GCs Velocity Dispersion and Mass estimation %%%%%%%%
\section{Recipe for a bias-corrected Velocity Dispersion and Mass estimators in Galaxy Clusters}
\label{sec:recipe}
Using DM only or hydro-dynamical cosmological simulations, \cite{evrard08},
\cite{munari13}, and \cite{saro13} characterised scaling relations between GCs
velocity dispersion of tracers, namely DM particles, sub-haloes and galaxies, and
$M_{200}$:
\begin{equation}
\frac{\sigma_{\rm 1D}}{\mathrm{km\ s^{-1}}} =
A\left[\frac{h(z)\ M_{200}}{10^{15}\ \mathrm{M}_{\odot}}\right]^{\alpha},
\label{eq:smg}
\end{equation}
where $\sigma_{\rm 1D} \equiv \sigma_{\rm 3D}/\sqrt{3}$, and the 3D velocity
dispersion, $\sigma_{\rm 3D}$, is calculated using all the DM particles or
galaxies within a sphere of radius $R_{200}$, using the biweight estimator
\citep{beers90}. However, DM particles, sub-haloes and galaxies lead to different
values of parameters $A$ and $\alpha$ \citep{munari13}. Moreover, owing to the
triaxiality of GCs and to the non-virialized state of some clusters, all the
constraints for $\alpha$ are slightly different from the value $\alpha=1/3$
derived from the virial theorem.

In order to use GCs for cosmological studies, it is crucial to obtain an
accurate, precise, and unbiased estimate for the velocity dispersion and,
consequently, for the cluster masses. Among other possibilities, this goal could
be achieved through spectroscopic follow-ups
\citep[e.g.,][]{allen11}. Nevertheless, owing to observational limits, in real
GC observations, it is very expensive to measure the line-of-sight velocity of
all cluster members. In the new era of large galaxy cluster samples, where we
often find galaxy cluster with a limited number of spectroscopic members
($N_{gal} \la 30$), it is important to characterise whether such a limited
number of galaxies might lead to biased estimates of the velocity dispersion
and/or the cluster mass.

The aim of this work is to characterise the statistical and physical biases both
for velocity dispersion and mass estimates in the regime of small number of
galaxies, and to provide a recipe to correct for them. As we explain below, the
four basic steps of our proposed recipe are:
\begin{enumerate}[label=\roman{*}., ref=(\roman{*})]
\item Evaluate the velocity dispersion of the cluster using an unbiased
  estimator;
\item Estimate the aperture radius and the mass fraction of the cluster members
  and correct for these sampling effects;
\item Estimate the fraction of interlopers that could contaminate the cluster
  members sample and correct the velocity dispersion;
\item Calculate the GC mass and correct for statistical biases introduced by the
  $\sigma$--$M$ relation.
\end{enumerate}
In the following sections we demonstrate that these four steps represent a good
way to estimate actual velocity dispersion and mass with samples containing low
numbers of galaxy members.

%%%%%%%%  SECTION 4: Bias in Velocity Dispersion estimation %%%%%%%%
\section{Statistical Bias and Variance for Velocity Dispersion estimators}
\label{sec:bias}

\subsection{Velocity dispersion estimators and notation}
\label{subsec:def}
\cite{beers90} presented a set of mean and scale estimators, and studied their
efficiency in the presence of deviations from a Gaussian-distribution. In this
paper, we decided to focus our attention on three of those estimators, namely
the standard deviation, the biweight and the gapper. The standard deviation,
\begin{equation}
\Sstd(\ngal) = \sqrt{\frac{1}{\ngal -1}\sum_{i=1}^{\ngal} (x_i - \mu)^2},
\label{eq:std_eq}
\end{equation}
is defined as the lowest variance scale estimator for a Gaussian
distribution. However, its dependence on $\mu$ (the mean of the distribution)
makes it a non-robust estimator.

The biweight scale estimator is a function of the sample median \citep{tukey58},
and it is defined as
\begin{equation}
S_{\rm bwt}(\ngal)  =
\left(\frac{\ngal^2}{\ngal-1}\right)^{1/2}\frac{\left[\Sigma_{|u_i|<1}\,(x_i-M)^2\,
    (1-u_i^2)^4\right]^{1/2}}{\left| \Sigma_{|u_i|<1}\, (1-u_i^2)\, (1-5u_i^2)
  \right|},
\label{eq:bwt_eq}
\end{equation}
where the $u_i$ quantities are given by
\begin{equation}
u_i=\frac{(x_i-M)}{a \times \mathrm{MAD}}
\label{eq:bwt_eq1}
\end{equation}
with $a=9.0$, and $\mathrm{MAD} = \mathrm{median} \,(|x_i-M|)$ are the tuning
constant and the median absolute deviation respectively. Finally, the gapper is
a robust estimator \citep{gapper} based on the gaps of an order statistics,
$x_i,x_{i+1},\dots,x_n$. It is defined as a weighted average of gaps:
\begin{equation}
S_{\rm gap}(\ngal) =\frac{\sqrt{\pi}}{\ngal \,(\ngal-1)}\sum_{i=1}^{\ngal-1} w_i \,g_i,
\label{eq:gap_eq}
\end{equation}
where the gaps are given by
\begin{equation}
g_i = x_{i+1} -x_i, \qquad i=1,\dots, \ngal -1
\label{eq:gap_eq1}
\end{equation}
and the (approximately Gaussian) weights are given by
\begin{equation}
w_i = i\,(\ngal -1).
\label{eq:gap_eq2}
\end{equation}
For a more detailed description of these estimators see \cite{beers90}. With the
notation introduced in equations~\ref{eq:std_eq}, \ref{eq:bwt_eq} and
\ref{eq:gap_eq}, throughout this paper we will refer in a generic way to any of
the three scale estimators as $\Sx(\ngal)$, being X$=$ ``std'', ``bwt'' or
``gap'', for each one of the three cases.

\subsection{Statistical bias and variance for the three estimators}

\begin{figure*}
\begin{center}
\includegraphics[width=0.49\textwidth]{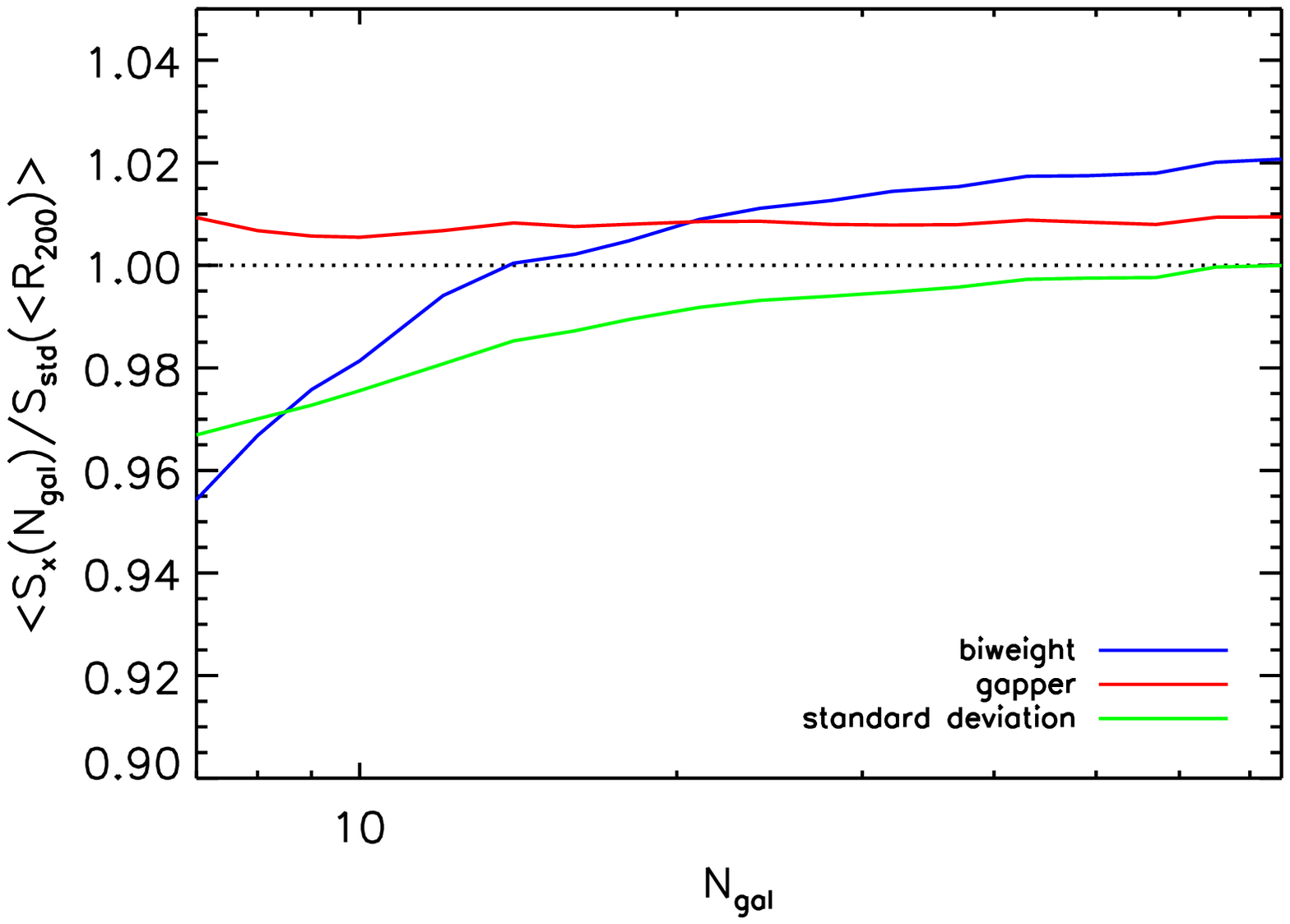}
\includegraphics[width=0.49\textwidth]{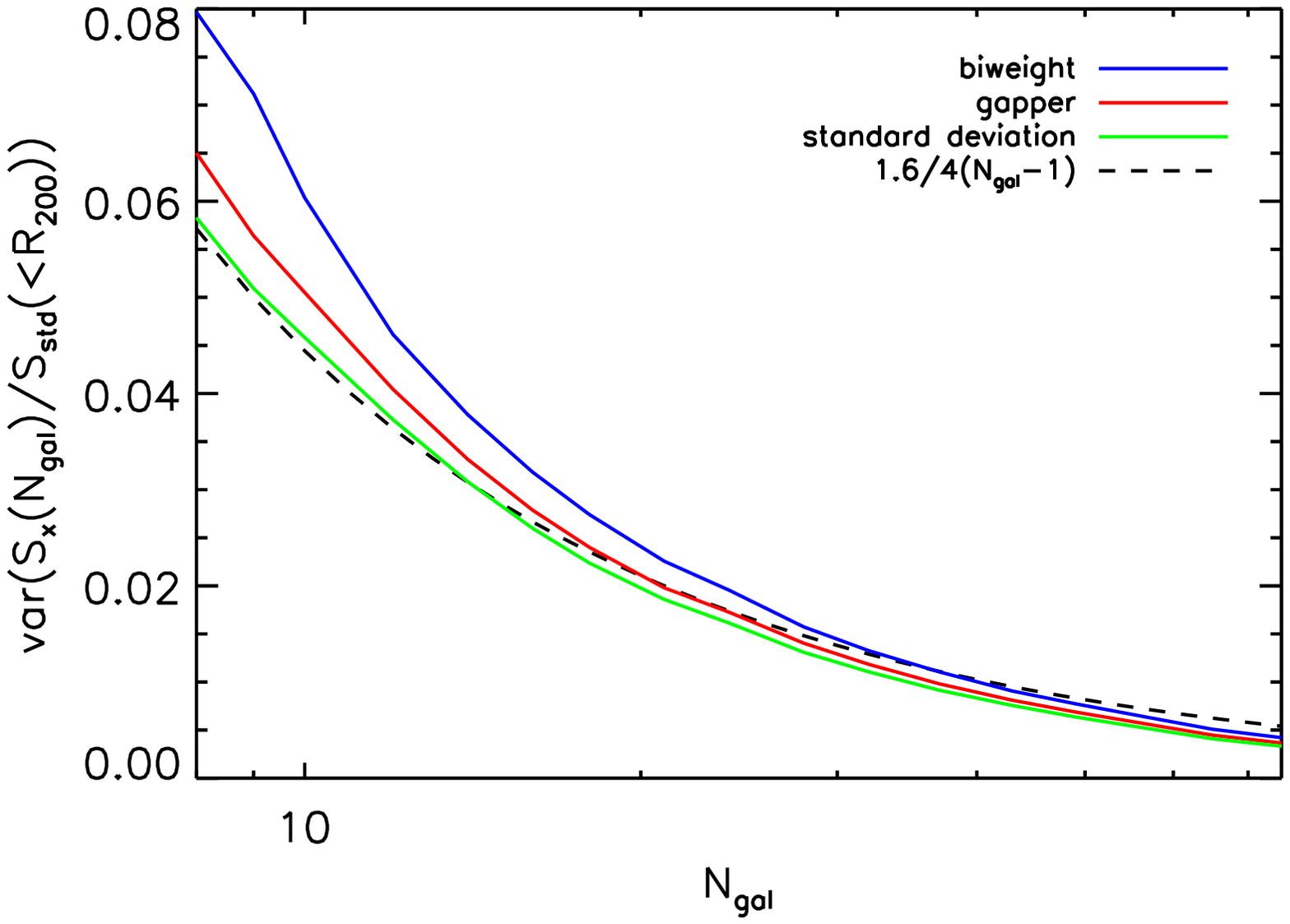}
\end{center}
\caption{The mean velocity dispersion $\Sx / \Sstd(<R_{200})$ as a function of
  the number of galaxies $\ngal$ (left panel) and its variance (right panel),
  for our sample of 73 simulated galaxy clusters. The dispersion $\Sx(\ngal)$ is
  calculated for the standard deviation (green line), biweight (blue line), and
  gapper (red line) estimators. }
  \label{fig:n_sigma}
\end{figure*}

Our aim in this work is to characterise the statistical behaviour of
the three aforementioned methods as a function of the number
of galaxies, by quantifying the possible bias of each technique
specifically in the small number of galaxies regime. As explained in
Sect.~\ref{sec:sim}, we use a set of 73 simulated GCs with 
redshifts $0.12 \leq z \leq 0.82$ and masses $2 \leq M_{200}/(10^{14}\,
\mathrm{M}_{\odot}) \leq 20$. Following the definition in \cite{munari13}, we
considered as galaxies only those DM subhaloes that contain a bounded stellar
structure with a mass $\geq 3 \times 10^{9}\, \mathrm{M}_{\odot}$.

We first characterise the distribution of velocities in our set of simulations.
Figure~\ref{fig:histos} shows the histogram of the radial velocities for the 73
GCs along the three main projection axes, of all cluster members
contained in a cylinder of projected radius $R_{200}$ along each
  selected axis. Even though
these global distributions are apparently close to 
a Gaussian, each one of the 73 individual GC distribution is not, due to the
present of substructures. A quantitative analysis shows that indeed there is a
deviation from gaussianity in the overall distributions. In particular,
  we have estimated the following dimensionless parameter
\begin{equation}
c \equiv \frac{\langle x_i^4\rangle-\langle x_i^2\rangle^2}{\langle
  x_i^2\rangle^2}, 
\label{eq:c_varvar}
\end{equation}
which is related to the fourth moment of the distribution. We would expect
$c=2$ for a perfect Gaussian sample. However, when evaluating the factor $c$ for
each of the 73 clusters, we find a mean value $c=1.6 \pm 0.4$ that implies a
departure from a Gaussian of the simulated GC velocity
distributions. As expected for relaxed clusters, the mean
  value of the $c$ parameter is found to be smaller than 2.
The quoted error of $0.4$ corresponds to the scatter of the
  $c$ parameter over the 73 simulated clusters.  As the statistical
  error in the determination of the $c$ parameter is significantly
  smaller than this value (on average, the number of galaxies within 
  $R_{200}$ for each cluster is 239, so naively we would expected a statistical error
  of the order of $1/\sqrt{239}=0.065$ for one cluster, and less than
  $0.01$ for the ensemble of 73 clusters), this large scatter is
  reflecting the intrinsic variety of clusters properties in our 
  simulations. We will use this $c$ parameter below, when estimating
the variance of the three estimators. 

% Bias
We now evaluate the bias and the variance of the three scale estimators
$\Sx(\ngal)$. To do this, we have explored 20 different values for $\ngal$,
between $\ngal =7$ and $\ngal =75$, logarithmically spaced to better analyze the
low-$\ngal$ tail. We have generated 2250 configurations by randomly selecting
galaxies projected in a circle of radius $R_{200}$, 750 times for each main axis
as line of sight and avoiding galaxy repetition. For each configuration, we
estimated $\Sx(\ngal)$ by repeating this procedure for each $\ngal$ and for each
galaxy cluster. The average values for $\Sx(\ngal)$ are obtained by averaging
the $73 \times 2250$ velocity dispersions normalised with respect to
$\Sstd(<R_{200})$, which represents the velocity dispersion of all the galaxies
in the simulation within a circle of projected radius $R_{200}$, and calculated
using the standard deviation estimator.
For completeness, we present in Table~\ref{table:rel_sigma_true} the ratio
of this relative bias when calculated with different
estimators. 

\begin{table}
\caption{Ratio of the relative bias $\Sx(<R_{200}) / S_{\rm Y}(<R_{200})$
  between two estimators X and Y, obtained with all the galaxies
in the simulation within a circle of projected radius $R_{200}$.}
\label{table:rel_sigma_true}
\centering
\begin{tabular}{c c c c} 
\hline\hline 
\noalign{\smallskip}
X / Y & $BWT$ & $GAP$ & $STD$ \\
\hline 
\noalign{\smallskip}
$BWT$ & $1.000$ & $1.013\pm0.009$ & $1.021\pm0.017$ \\ 
$GAP$ & $0.987\pm0.008$ & $1.000$ & $1.008\pm0.010$ \\
$STD$ & $0.980\pm0.017$ & $0.992\pm0.010$ & $1.000$ \\
\noalign{\smallskip}
\hline 
\end{tabular}
\end{table}

In the left panel of Fig.~\ref{fig:n_sigma} we show how each estimator is able
to recover the velocity dispersion, when compared to the standard deviation of
the full sample $\Sstd(< R_{200})$. By construction, for high $\ngal$ (i.e.,
when using all galaxies in the simulation within $R_{200}$), the standard
deviation estimator $\Sstd(\ngal) / \Sstd(<R_{200})$ tends to one, while the
other two estimators recover the asymptotic value given in
Table~\ref{table:rel_sigma_true}.

In the low-$\ngal$ regime, all estimators are biased. The gapper (red line)
returns an almost constant estimate of the velocity dispersion at any $\ngal$,
but that average value is slightly biased with respect to the true variance
($1.008 \pm 0.010$, as shown in Table~\ref{table:rel_sigma_true}). The biweight
shows a stronger dependence on the number of elements used for the estimation,
specially in the low-$\ngal$ regime. In fact, for $\ngal$ smaller than 30, it
underestimates the true dispersion by up to $4$\,\% at $\ngal = 10$. A very
similar behaviour is shown by the standard deviation estimator. We note that in
this later case, the dependence on $\ngal$ can be theoretically predicted, as
showed in Appendix~\ref{apendix:ap1}, giving the analytic form $1 -1/(4 (\ngal
-1))$.
Based on this dependence on $\ngal$, we have obtained a numerical fit to those
curves in the left panel of Figure~\ref{fig:n_sigma}, using the following
parametric equation:
\begin{equation}
1 - \left(\left(\frac{D}{(N_{\rm gal}-1)}\right)^\beta+B\right).
\label{eq:parfit_sigma}
\end{equation}
Table~\ref{table:param_un_sim} shows the best-fit values for the parameters $D$,
$\beta$ and $B$, for each one of the three estimators (biweight, gapper and
standard deviation).

Another crucial aspect for choosing a $\sigma_v$ estimator is its variance. We
would expect the standard deviation to be the lowest variance estimator for a
Gaussian distribution. We illustrate this in Appendix~\ref{apendix:ap2}, where
we also show the behaviour of all three estimators in the same limit of Gaussian
velocity distributions.
For the more realistic case given by our set of numerical simulations, we
confirm that this is also the case. The right panel of Fig.~\ref{fig:n_sigma}
shows the variance of the three estimators, $Var( \Sx(\ngal) )$, and it shows
that the standard deviation has still the lowest variance. Moreover, we can
compare this measured variance with the optimal one expected for the theoretical
behaviour for a homogeneous population given by\footnote{To derive this equation, we have
  used the definition of the $c$ parameter, and that the variance of
  the variance of a centred random variable $x$ can be computed as
  $(<x^4> - <x^2>^2)/N$, being $N$ the number of data samples. }
\begin{equation}
Var( \Sstd(\ngal) / \Sstd(< R_{200}) ) = \frac{c}{4 (\ngal -1)},\\
\label{eq:v_sig_teoric_2}
\end{equation}
where the parameter $c$ was defined in
equation~\ref{eq:c_varvar}. We find that
the variance of the standard deviation is indeed very close to the optimal one,
as well as the  variance of the gapper. For low $\ngal$
values ($\la 20$), the variance of the biweight estimator is significantly
worse. Numerical fits to the dependence of the variance as a
  function of $\ngal$ are given in Appendix~\ref{apendix:ap4}.

Using either the parametric fitting given in equation~\ref{eq:parfit_sigma}, or
the numerical values from the left panel of Figure~\ref{fig:n_sigma}, we can now
construct unbiased velocity dispersion estimators, by explicitly correcting for
that statistical bias. We will use the primed notation $\Sxprime(\ngal)$ when
referring to these ``corrected'' estimators, which will be given by
\begin{align}
\label{eq:un_S}
\Sxprime(\ngal) \equiv \Sx(\ngal) \left(1 - \left(\left(\frac{D}{(\ngal
  -1)}\right)^\beta+B\right)\right)^{-1} \\
\nonumber
\approx \Sx(\ngal) \left(1 + \left(\left(\frac{D}{(\ngal -1)}\right)^\beta+B\right)\right).
\end{align}
and where the approximation in the second line uses the fact that the correction
term is small compared to unity.

\begin{figure*}
\begin{center}
\includegraphics[width=0.49\textwidth]{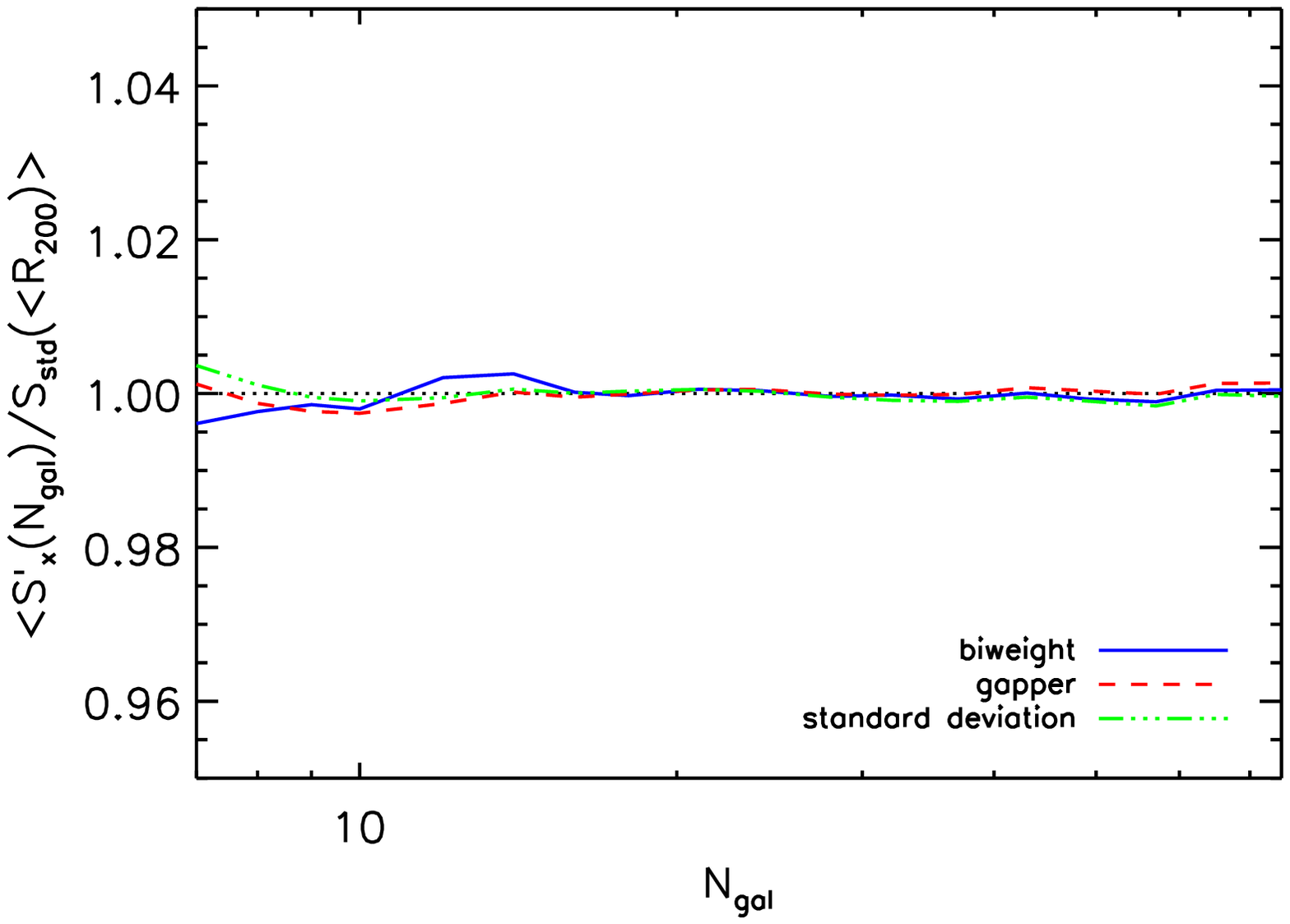}
\includegraphics[width=0.49\textwidth]{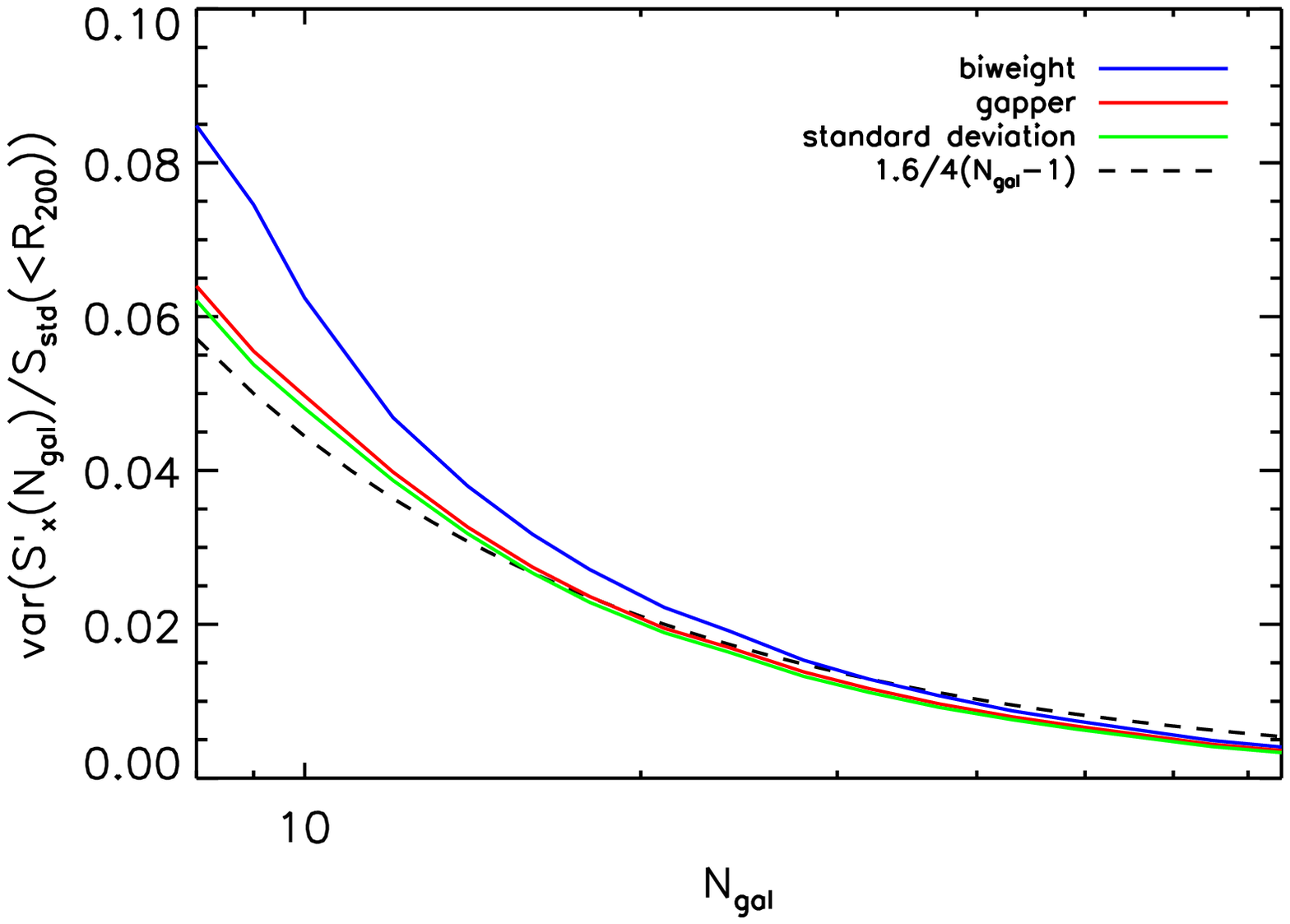}
\end{center}
\caption{Mean (left panel) and variance (right panel) of the corrected
  estimators $\Sxprime(\ngal)/S'_{\rm std}(<R_{200})$, as a function of the
  number of galaxies $\ngal$ in our sample of 73 simulated galaxy clusters, for
  the standard deviation (green line), biweight (blue line), and gapper (red
  line). }
  \label{fig:n_sigma2}
\end{figure*}

\begin{table}
\footnotesize
\caption{Best-fit parameters to be used in the parametric function given in
  equation~\ref{eq:parfit_sigma}, describing the bias of the three
  estimators. See text for details.  }
\label{table:param_un_sim} 
\centering 
\begin{tabular}{c c c c} 
\hline\hline 
\noalign{\smallskip}
\, & $BWT$ & $GAP$ & $STD$ \\
\hline
\noalign{\smallskip}
%$D$ & $0.72\pm0.03$ & $0$ & $0.25$\\ 
%$B$ & $-0.0225\pm0.0002$ & $-0.0080\pm0.0002$ & $-0.0037\pm0.0003$\\
%$\beta$ & $1.28\pm0.03$ & $ 1 $ & $ 1 $ \\
$D$ & $1$ & $0$ & $0.25$\\ 
$B$ & $-0.0124\pm0.0009$ & $-0.0018\pm0.0005$ & $-0.0016\pm0.0005$\\
$\beta$ & $1.43\pm0.01$ & $ 1 $ & $ 1 $ \\

\noalign{\smallskip}
\hline 
\end{tabular}
\end{table}

Figure~\ref{fig:n_sigma2} is equivalent to the Fig.~\ref{fig:n_sigma}, but now
computed for the set of corrected estimators defined in equation~\ref{eq:un_S}.
By construction, the new $\Sxprime(\ngal)$ estimators are now unbiased (left
panel), and their variance (right panel) have increased only by an small amount.
As for the case of unprimed estimators, the corrected standard deviation is
still the minimum variance estimator, although the three of them present very
similar values for $\ngal \ga 30$.
For this reason, we decided to use $S'_{\rm std}(\ngal)$ as the reference
estimator in the following sections, although we could in principle use any of
the three estimators.

%%%%%%%%  SECTION 5: Bias from interlopers contamination %%%%%%%%
\section{Bias from interlopers contamination}
\label{subsec:bias_inter}

Galaxy clusters are not isolated structures in the Universe. This fact,
  together with the inevitable confusion associated to redshift-space
  measurements, implies that any spectroscopic sample of potential cluster
  members could be in principle contaminated. This population of pseudo cluster
  members, called ``interlopers'', modifies the velocity distribution and
  therefore affects the estimation of the velocity dispersion
  \citep[e.g.][]{Wojtak07, Wojtak18, Pratt19}.
Using numerical simulations of the entire visual cone,
  \cite{mamon10} showed that the fraction of galaxies outside the virial sphere
  that appear on sky projected within the virial radius could reach up to $\sim
  27\,\%$, making the interlopers a potentially important source of error for an unbiased
  determination of the underlying velocity distribution. As shown below, the overall error due to interlopers is indeed
  similar or slightly larger than the statistical and physical biases
  discussed in this paper.
%\citep[e.g.][]{Wojtak07, Wojtak18, Pratt19}

According to the definition of interlopers given in \cite{Pratt19}, it
  is useful to consider this population as the sum of two different types of
  objects: (i) galaxies gravitationally bounded to the clusters that are far
  from the cluster centre, but due to projection effects appear within a
  projected circle of a smaller radius (hereafter ``type 1'' interlopers); and (ii)
  background/foreground galaxies with similar redshifts to that of the cluster,
  but belonging to the large scale structure that surrounds the cluster itself
  (hereafter ``type 2'' interlopers). Note that, in our particular case of zoomed
  simulations, they are only including type 1 interlopers.

Providing a general recipe to correct the velocity dispersion bias of a
  given estimator due to the presence of interlopers is not possible, as in
  general the fraction of those objects will depend not only on $\ngal$, but
  also on the particular criteria adopted for assigning cluster membership to
  galaxies observed in the cluster field, as well as the type of interlopers.
  Moreover, there are multiple methods in the literature for identifying
  interlopers, usually linked to specific cluster mass reconstruction methods. A
  very complete list can be found in \citet{Wojtak18} and references therein.
  Unfortunately, none of those methods are capable of completely removing
  all contaminants \citep{Wojtak18}, and moreover, these techniques
  have good results when applied to large galaxy samples (hundreds of
  members), but they are usually less effective when applied to
  smaller samples (tenths of members), as in the case of the caustic
  method \citep{diaferio99}.

Here we limit our discussion to one particular member
  selection method, named the ``sigma clipping'', and we illustrate the
  procedure to carry out the correction of the velocity estimation for the two
  types of interlopers. We emphasise that, in a general case, specific
  simulations will be required to quantify the bias associated to each
  particular method. The sigma clipping \citep{YahilVidal1977} is one of the
  most used techniques for removing interlopers. This method clips galaxies
  whose radial velocity is above a certain threshold, being
  particularly effective in the external regions of the clusters.

\subsection{Type 1 interlopers}
\label{subsec:bias_inter_ref}

We first estimate the impact of type 1 interlopers in our
  simulations. It is important to emphasise here that, throughout this
  paper, all our velocity dispersion quantities are computed
  using the galaxies contained in a cylinder of projected radius
  $R_{200}$. Thus, by construction, they will be affected by type 1
  interlopers. In order to transform them into a velocity
  dispersion computed within a sphere of radius $R_{200}$, and
  therefore, free of type 1 interlopers, the average conversion
  factors that we find in our simulated sample are $0.990$, 
  $0.981$ and $0.985$ for the biweight, gapper and standard deviation
  estimators, respectively.  In summary, as a consequence of the
  presence of (type 1)
  interlopers, the velocity dispersion is overestimated between $1\%$ and
  $2\%$. This bias is relatively small, and comparable to the
  statistical biases discussed in the previous sections. In
  principle, it has to be corrected in the corresponding $\Sxprime$ 
  estimators when transforming from velocity dispersion into masses, 
  but only if the adopted $\sigma$--$M$ scaling relation from simulations did already
  account for the effect of type 1 interlopers.

\begin{figure}
\begin{center}
\includegraphics[width=0.49\textwidth]{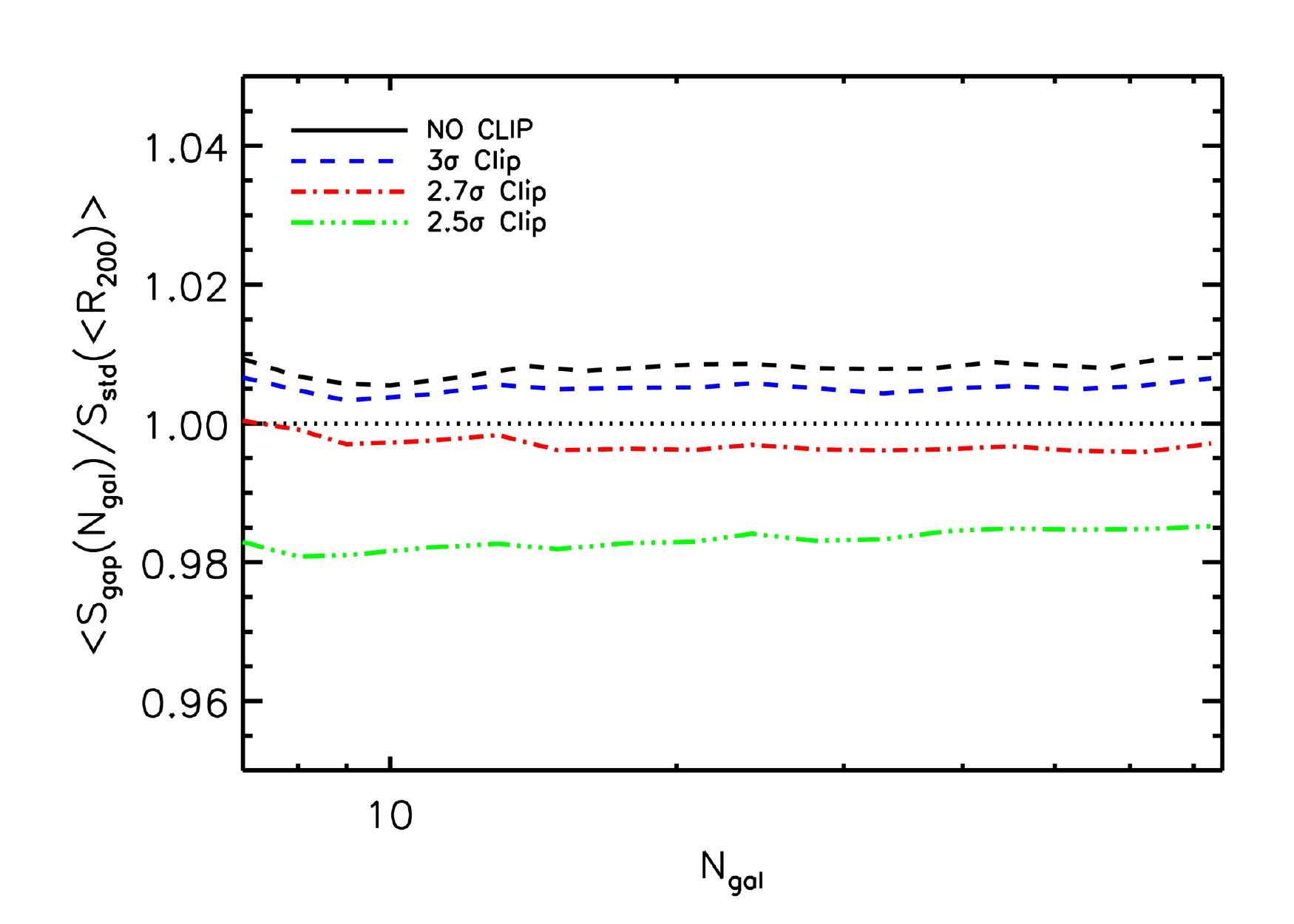}
\end{center}
\caption{Effect of type 1 interlopers on the velocity estimates, as a function of
  the number of galaxies, for the gapper estimator. The velocity
  dispersion $S_{\rm gap}(\ngal)$ is computed first using the full
  galaxy sample (black solid line, equivalent to the red line in
  Figure~\ref{fig:n_sigma}), and then it is compared to the estimates  
  after clipping the galaxy sample at $3$, $2.7$ and $2.5$ sigmas in
  the velocity space.  }
  \label{fig:interlopers1}
\end{figure}

We further explore the possible correction of this bias using
  the sigma clipping method. Although this method might be effective in removing
  interlopers,  cutting the tail of a distribution will necessarily introduce a bias in the estimation
  of the velocity dispersion. To quantify this effect, we tested four grades of
  clipping (no clip, $3\sigma$, $2.7\sigma$, $2.5\sigma$) in
  Figure~\ref{fig:interlopers1} for one of the estimators. As expected, the higher the clip, the
  lower is the variance recovered by the estimator. However, it is
  interesting to note that this new bias partially alleviate
  the effect introduced by the type 1 interloper contamination. As in \citet{mamon10}, we also find that both effects are compensated at
  around $2.7\sigma$, if we use the gapper to estimate the dispersion.  However, we note
  that using the other two estimators, the clipping that compensate the effect of
  the interlopers is different. We find that a  $2.5\sigma$ clipping for the biweight, and a $3\sigma$ clipping for the
  standard deviation compensates the effect of type-1 interlopers.

\subsection{Type 2 interlopers}
\label{subsec:bias_inter_T2}

\begin{figure}
\begin{center}
\includegraphics[width=0.49\textwidth]{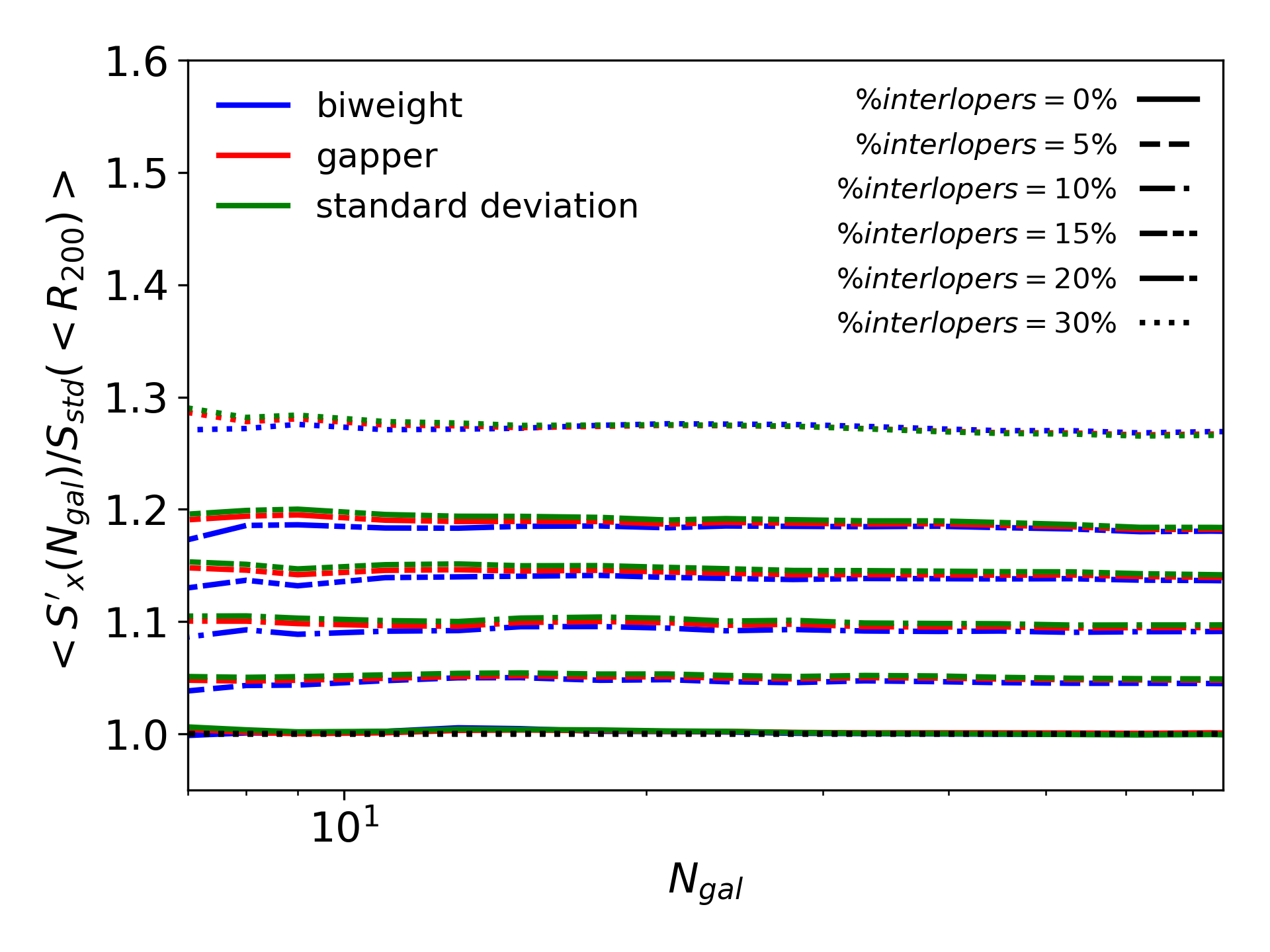}
\end{center}
\caption{Combined effect of type 1 and 2 interlopers on $\Sxprime(\ngal) / S_{\rm std}(<R_{200})$, as a
  function of the number of galaxies. The velocity dispersion,
  $\Sxprime(\ngal)$, is first computed using the full
  galaxy sample (equivalent to the three colour lines shown in the
  left pannel of Figure~\ref{fig:n_sigma2}). We also evaluated the response of biweight (blue), standard
  deviation (green), and gapper (red lines) using samples contaminated
  by a certain fraction of type 2 interlopers, as described in the
  text. By construction, our simulations also include the type 1
  interlopers, as the velocity dispersion is estimated in the
  cylinder. }
\label{fig:interlopers2}
\end{figure}
As explained above, here we use zoomed hydrodynamical
simulations from a parent one, where the region of clusters are re-simulated at higher
resolution. Although this technique is very useful to explore the
appropriate mass range to form stars and galaxies, it has the
disadvantage of re-simulating only a finite region around the centre
of the cluster (in the case of study $\sim 5 R_{200}$). Thus, all galaxies in our simulated catalogues are
bounded to the cluster, and following our definition, they correspond
to type 1 interlopers only.

To understand the effect introduced by type 2 interlopers, we have
repeated the procedure described in the previous sub-section, but this time
replacing some galaxies from the actual cluster distribution with
random velocity values drawn from a uniform distribution in the
velocity interval $[-2.7, 2.7] \times \Sstd(< R_{200})$. This uniform
distribution is intended to mimic a field of background and foreground
galaxies, in an extreme case of a velocity distribution which is completely different from that of the galaxies.  
Figure~\ref{fig:interlopers2} presents the results obtained for five different fraction of interlopers:
5\,\% (dashed lines), 10\,\% (dot-dashed lines), 15\,\%(three dot-dashed lines), 20\,\%
(long dashed lines), and 30\,\% (dotted lines). As expected, the
inclusion of those type 2 interlopers produces a positive bias in the velocity dispersion, which at first order is found to be directly 
proportional to the relative fraction of interlopers. It is also
noteworthy that all the three estimators are similarly affected by
this ``type 2'' interloper contamination.

Although Fig.~\ref{fig:interlopers2} shows a broad range of values for the fraction of type
  2 interlopers, in real objects we would expect this number to be in
  the range of 5 to 10\,\% within a virial radius \citep{saro13}, being
  the fraction of type 1 objects significantly larger in number. If this is the case, then the effect of type 2 interlopers
  in the extreme case considered here will be at most 10 per cent. 
This is consistent with other results in the literature, which indeed present smaller values. For example,
\citet{mamon10} showed that the the total fraction of interlopers in their simulations (including both types 1 and 2) is $\sim 27\%$, while
their impact on the velocity estimation at $R_{200}$ is of the order of 2\,\%. Moreover, this effect is basically cancelled out in their final 
estimation of the velocity dispersion within $R_{200}$ when using the $2.7\sigma$ clipping, thus suggesting that the fraction of type 2 interlopers
with a very different velocity distribution to the one of the true members is rather small. On the other hand, in our simulated cluster sample we find a median fraction of type 1
interlopers of $\sim 29\%$, which is consistent with the value of \citet{mamon10}.

It is also important to note the strong dependence of type 2
interlopers with the sample aperture \citep{mamon10, saro13}, increasing rapidly
beyond $R_{200}$. In practice, this makes the interloper contamination the most
damaging effect for obtaining an unbiased velocity estimation for a
single cluster for radii much larger than $R_{200}$.

Finally, we note that in real observations, the fraction of
interlopers, and particularly type 2,  will depend closely on the
observational strategy and the particular algorithms and procedures
used for member selection. Therefore, the effective number of contaminants
cannot be estimated precisely with a general recipe. Studies such as
\cite{mamon10, saro13} are necessary to statistically quantify their
abundance in each particular observing strategy.
As we are focused here in providing a general recipe for correcting the
statistical and/or physical bias associated to the velocity estimators, we will
not discuss further this effect. But we emphasise that for a reliable velocity
estimation, the bias due to interlopers has to be taken into account and
corrected specifically for each particular survey.

%%%%%%%%  SECTION 6: Physical Biases on velocity dispersion estimation %%%%%%%%
\section{Physical Biases on velocity dispersion estimators}
\label{sec:bias2}

In the ideal case in which we can choose an uniformly selected sample of true
cluster members inside $R_{200}$, the corrected set of scale estimators
presented above will provide an unbiased estimation of the velocity dispersion
of the cluster. However, observational strategies and technical limitations
prevent us from reaching the ideal case. In this section we study two possible
ways in which a particular selection of cluster members might produce biased
velocity dispersion estimates.

%POR AQUI

\subsection{Effects due to the selected fraction of massive galaxies}
\label{subsec:bias_mass}
For a fixed integration time, the telescope aperture limits the detection
magnitude and prevents us from detecting faint objects. In other cases, the
technical requirements of spectrographs make it impossible to sample the cluster
members adequately for arbitrary low brightness values. So, in practice,
line-of-sight velocity samples contain only a fraction of GC members, generally
the brightest objects in the GC, which are also the most massive. This fraction
of objects is particularly small for high redshift GCs. In this subsection, we
investigate if there is an induced bias due to this mass segregation. 

In order to simulate this effect, we mimicked observational conditions by
selecting three percentages of all visible galaxies in the simulation,
i.e. 50\,\%, 33\,\%, and 25\,\%, by sorting the cluster members by mass and
dividing the sample in 2, 3, and 4 mass bins, starting from the most massive
object. For each case, as explained in Sect.~\ref{sec:bias}, we averaged 2250
configurations (750 for each axis, $x$, $y$, $z$), considering numbers of
galaxies between 8 and 75, avoiding galaxy repetition and evaluating the
dispersion with the biweight, gapper, and standard deviation methods.

\begin{figure*}
\begin{center}
\includegraphics[width=0.49\textwidth]{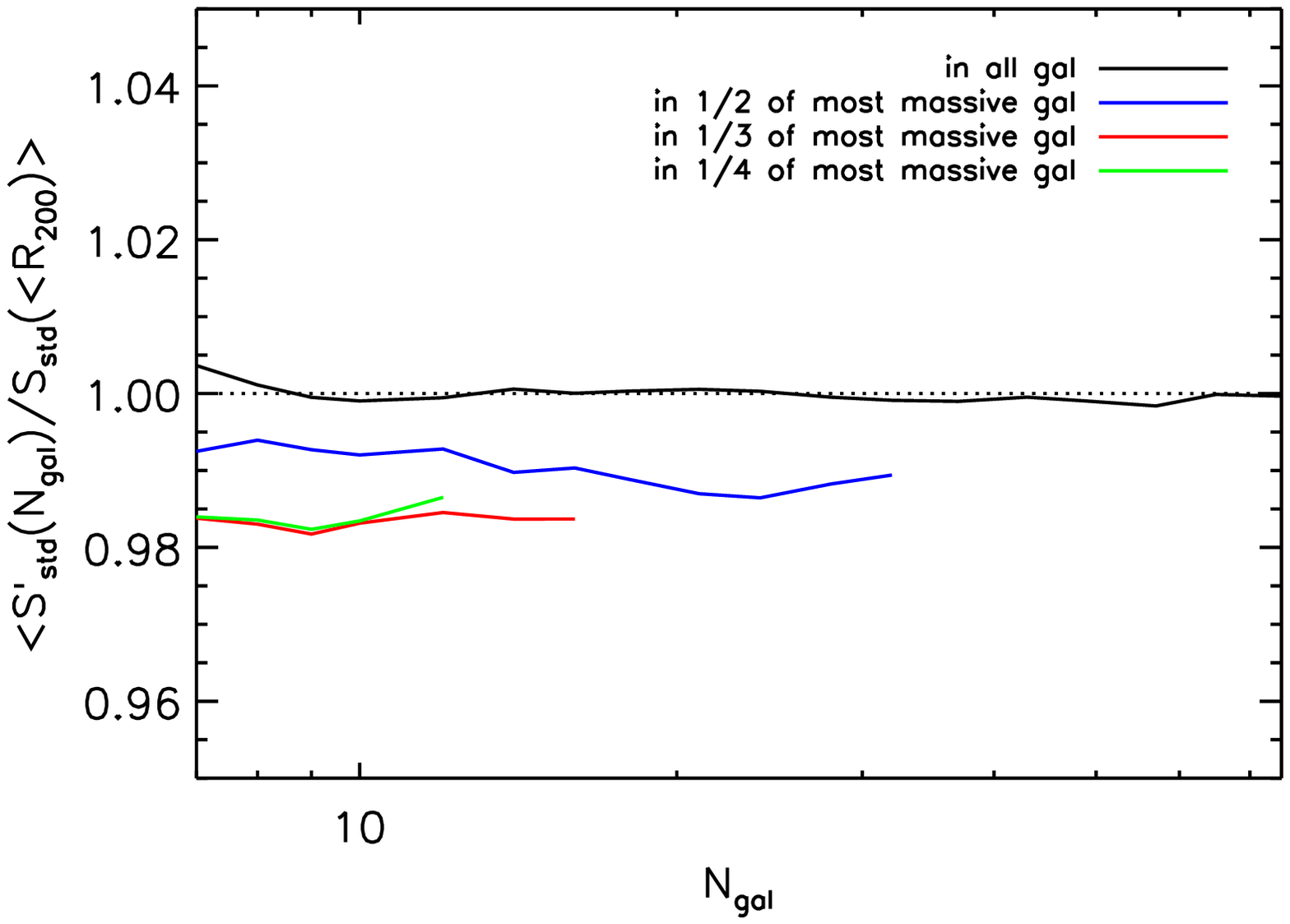}
\includegraphics[width=0.49\textwidth]{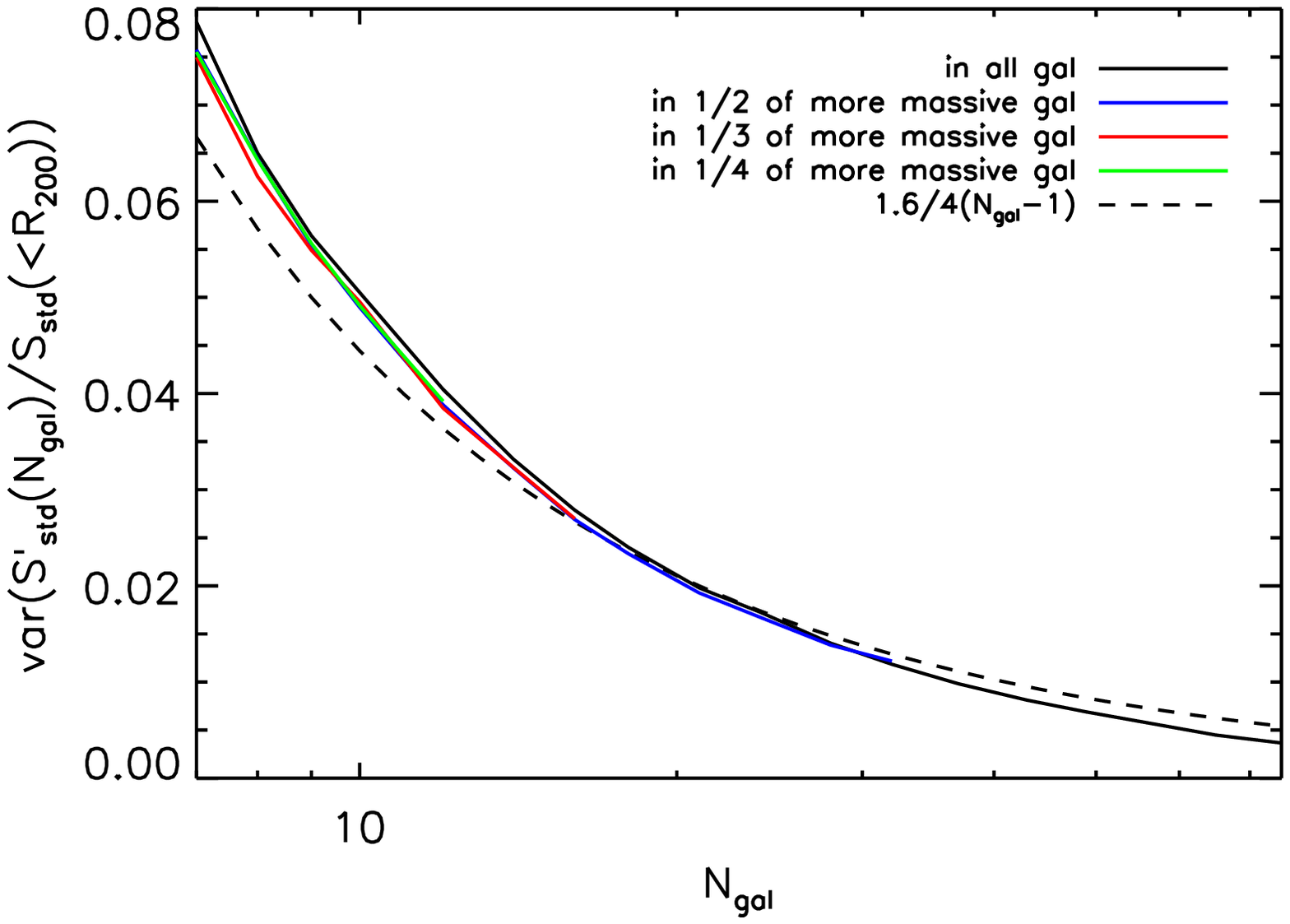}
\end{center}
\caption{Left panel: Mean (bias) of $S'_{\rm std}(\ngal)/ \Sstd(<R_{200})$ as a
  function of the number of galaxies $\ngal$, calculated by choosing galaxies
  within $100\%$ (black solid line), $1/2$ (blue solid line), $1/3$ (red solid
  line), and $1/4$ (green solid line) of the complete cluster member
  samples. Right panel: Variance of $S'_{\rm std}(\ngal)/ \Sstd(<R_{200})$ as a
  function of the number of galaxies $\ngal$. The dashed line represents the
  theoretical expectation for the variance of the dispersion according to
  equation~\ref{eq:v_sig_teoric_2}. }
  \label{fig:n_sigma_ahtqo}
\end{figure*}

Figure~\ref{fig:n_sigma_ahtqo} (left panel) shows $S'_{\rm
  std}(\ngal)/\Sstd(<R_{200})$ as function of $\ngal$ calculated with the
corrected standard deviation estimator, and using galaxies picked up from
100\,\% (black line), $1/2$ (blue line), $1/3$ (red line) and $1/4$ (green line)
of the complete cluster member samples. We see how a bias appears, with a
nonlinear dependence with the fraction of massive galaxies considered in each
case, but almost insensitive to the $\ngal$ parameter. This means that the
velocity dispersion is sensitive to the fraction of massive galaxies used to
estimate it. In particular, taking into account only the most massive galaxies
of the clusters ($1/4$ of the sample), one would find a velocity dispersion that
could be underestimated up to 2 per cent. We can interpret this velocity bias in
terms of a physical mechanism, the dynamical friction \citep{chandrasekhar43},
which mostly affects the most massive galaxies, so that the velocity dispersion
is lower with respect to that obtained using objects randomly selected from the
complete galaxy sample. Table~\ref{table:frac_bias} shows the average bias of
the primed estimators $\Sxprime$, calculated with respect to the full set of
cluster members within $R_{200}$ (i.e., $\Sstd(<R_{200})$), for each fraction in
exam, and for the three estimators (biweight, gapper, and standard
deviation). As this physical bias is almost independent on $\ngal$, we could in
principle use directly those values to produce a new corrected (unbiased)
estimator.

\begin{table}
\caption{Relative bias of the primed estimators due to the selected fraction of
  massive galaxies. We evaluate it as the average $\left<
  \Sxprime/\Sstd(<R_{200}) \right>$ for all possible $\ngal$ values. }
\label{table:frac_bias} 
\centering 
\begin{tabular}{c c c c} 
\hline\hline 
\noalign{\smallskip}
Fraction & $BWT$ & $GAP$ & $STD$ \\
\hline 
\noalign{\smallskip}
$1$ & $1.013$ & $1.008$ & $1$ \\ 
$1/2$ & $1.004$ & $0.999$ & $0.990$\\
$1/3$ & $0.996$ & $0.992$ & $0.981$\\
$1/4$ & $0.992$ & $0.993$ & $0.982$\\
\noalign{\smallskip}
\hline 
\end{tabular}
\end{table}
In the right panel of Figure~\ref{fig:n_sigma_ahtqo} we show the variance of
$S'_{\rm std}(\ngal)/\Sstd(<R_{200})$. The fraction of massive galaxies does not
significantly affect the dispersion estimator variance.

\subsection{Effect of aperture sub-sampling}
\label{subsec:bias3}
\begin{figure}
\begin{center}
\includegraphics[width=0.49\textwidth]{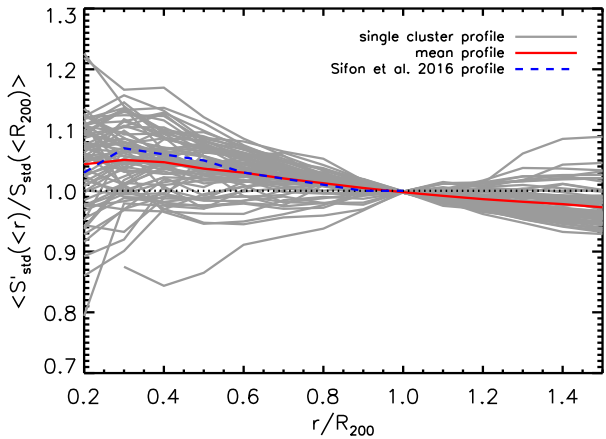}
\end{center}
\caption{Average velocity dispersion profile within a given enclosed radius $r$,
  $<S'_{\rm std}(<r)/\Sstd(<R_{200})>$, normalised to $R_{200}$. The red line
  represents the mean at each radius of the individual 73 simulated GC profiles
  (grey lines). The numerical values are given in Table~\ref{tab:rad_bias}. The
  dashed blue line represents the \cite{sifon16} profile, which is almost
  coincident with our derived profile.  }
  \label{fig:r_sigma_ahtqo}
\end{figure}

All the analyses presented above include galaxies from the complete sample of
cluster members, or a fraction of them, but the sample always being selected
within $R_{200}$. However, there is already evidence in the literature that the
velocity dispersion estimate needs to be corrected if galaxies are not sampled
out to the cluster's virial radius \citep[e.g.,][]{mamon10,sifon16}.
In this subsection, we want to see how the selection region affects the
$\sigma_v$ estimate in our simulations, by characterising the physical bias
introduced when evaluating the velocity dispersion enclosed in a radius $r$ from
the galaxy cluster center. In particular, we compute the velocity dispersion
using all the galaxies inside a cylinder of variable radius $0.2\leq r/R_{200}
\leq 1.5$. In addition, we average over all 73 simulated GCs and construct the
$\left< \Sxprime(<r)/\Sstd(<R_{200})\right>$ as a function of the $r /R_{200}$
profile (red line Fig.~\ref{fig:r_sigma_ahtqo}). The corresponding numerical
values are given in Table~\ref{tab:rad_bias}.  The velocity dispersion is (on
average) overestimated where the region explored by the spectroscopic sample is
smaller than $R_{200}$. The results are consistent with those obtained by
\cite{sifon16} when using the biweight estimator for both $\Sxprime(<r)$ and
$\Sxprime(<R_{200})$ (see Figure~4 and Table~3 in that paper).

We finally evaluate the combined effect of this aperture sub-sampling and the
selection effect of a fraction of massive galaxies discussed in the previous
subsection. We performed the same analysis described above for eight fractions
of $R_{200}$, i.e. $r/R_{200}=0.2$, $0.3$, $0.4$, $0.5$, $0.8$, $1.0$, $1.2$ and $1.5$, and
considering also different fractions of massive members.  Unfortunately, we were
unable to calculate the dispersion for the smaller fractions for radii less than
half $R_{200}$ for our simulation set because not all the simulated clusters
contain at least seven galaxies in these inner regions. But for larger radii
($r\ga R_{200}$), we find that the bias caused by the used fraction of massive
galaxies remains almost constant at all radii. Therefore, we can apply the
correction factors presented in Table~\ref{table:frac_bias}, in combination with
the radial correction profile shown in Figure~\ref{fig:r_sigma_ahtqo} and in
Table~\ref{tab:rad_bias}, to correct simultaneously for both effects.

\begin{table}
\caption{ Average velocity dispersion profile within a given enclosed radius
  $r$, $<S'_{\rm std}(<r)/\Sstd(<R_{200})>$, normalised to $R_{200}$. Values
  computed from the simulations. Uncertainties are the standard deviation.  }
\label{tab:rad_bias}
\centering
\begin{tabular}{c c} 
\hline\hline 
\noalign{\smallskip}
\, $r/R_{200}$ & $S'_{\rm std}(<r)/\Sstd(<R_{200})$ \\
\hline 
\noalign{\smallskip}
$0.2$ & $1.044 \pm 0.128$ \\ 
$0.3$ & $1.051 \pm 0.106$ \\ 
$0.4$ & $1.047 \pm 0.089$ \\ 
$0.5$ & $1.036 \pm 0.071$ \\ 
$0.6$ & $1.030 \pm 0.053$ \\ 
$0.7$ & $1.020 \pm 0.039$ \\ 
$0.8$ & $1.012 \pm 0.026$ \\ 
$0.9$ & $1.005 \pm 0.015$ \\ 
$1.0$ & $0.998 \pm 0.001$ \\ 
$1.1$ & $0.992 \pm 0.016$ \\ 
$1.2$ & $0.986 \pm 0.024$ \\ 
$1.3$ & $0.982 \pm 0.034$ \\ 
$1.4$ & $0.978 \pm 0.042$ \\ 
$1.5$ & $0.973 \pm 0.045$ \\
\noalign{\smallskip}
\hline 
\end{tabular}
\end{table}

%%%%%%%%  SECTION 6: Statistical Bias in $M_{200}$ estimation %%%%%%%%
\section{Bias in the mass estimation}

\subsection{Statistical bias in the estimation of $M_{200}$}
\label{sec:s_mass_bias}

In the previous sections we have studied how velocity dispersion estimators can
be affected by different statistical and physical factors, and we have
quantified the expected bias in those cases. In this section, we now show how
mass estimators are also affected by the same effects.

\begin{figure*}
\begin{center}
\includegraphics[width=0.49\textwidth]{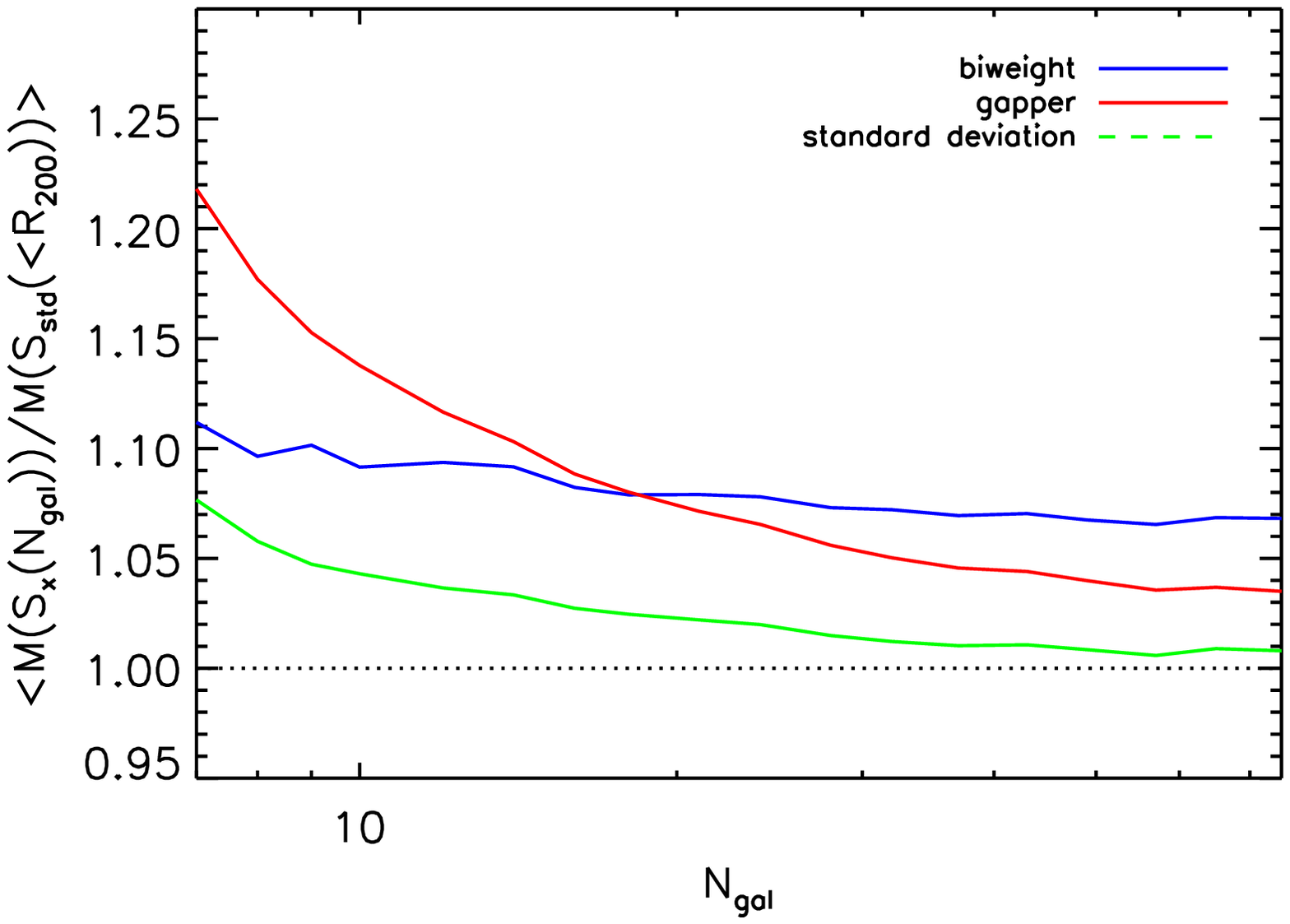}
\includegraphics[width=0.49\textwidth]{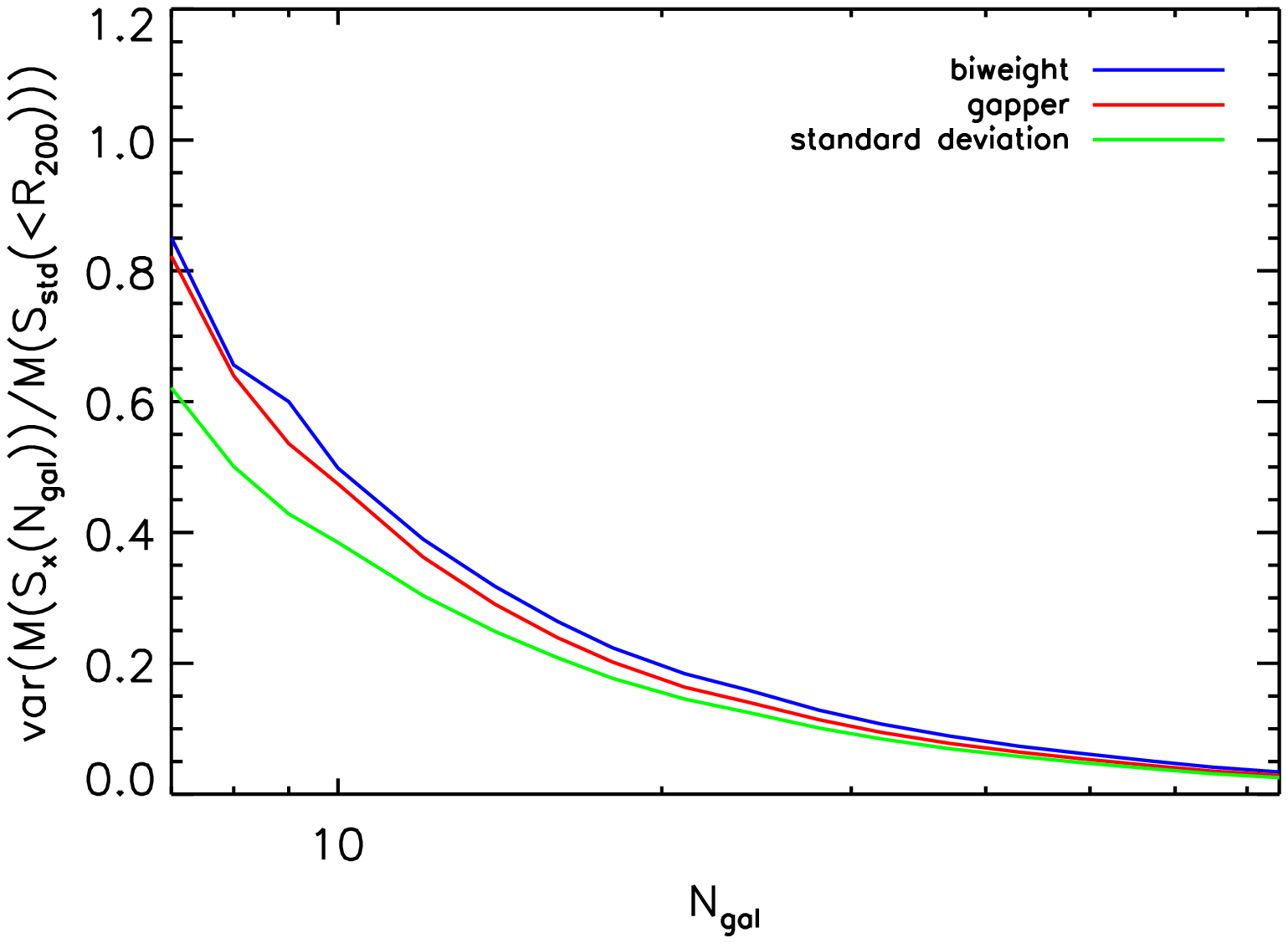}
\includegraphics[width=0.49\textwidth]{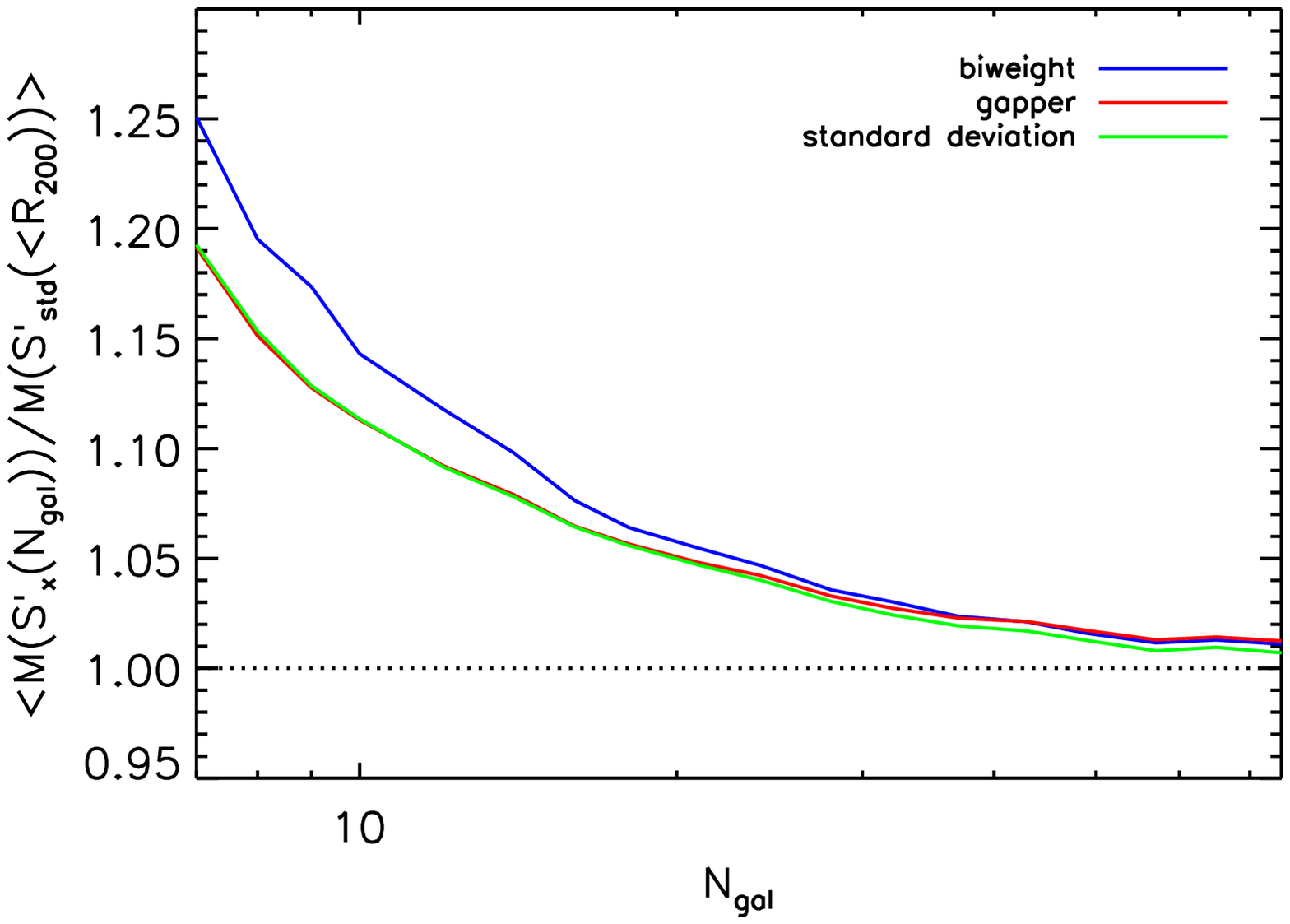}
\includegraphics[width=0.49\textwidth]{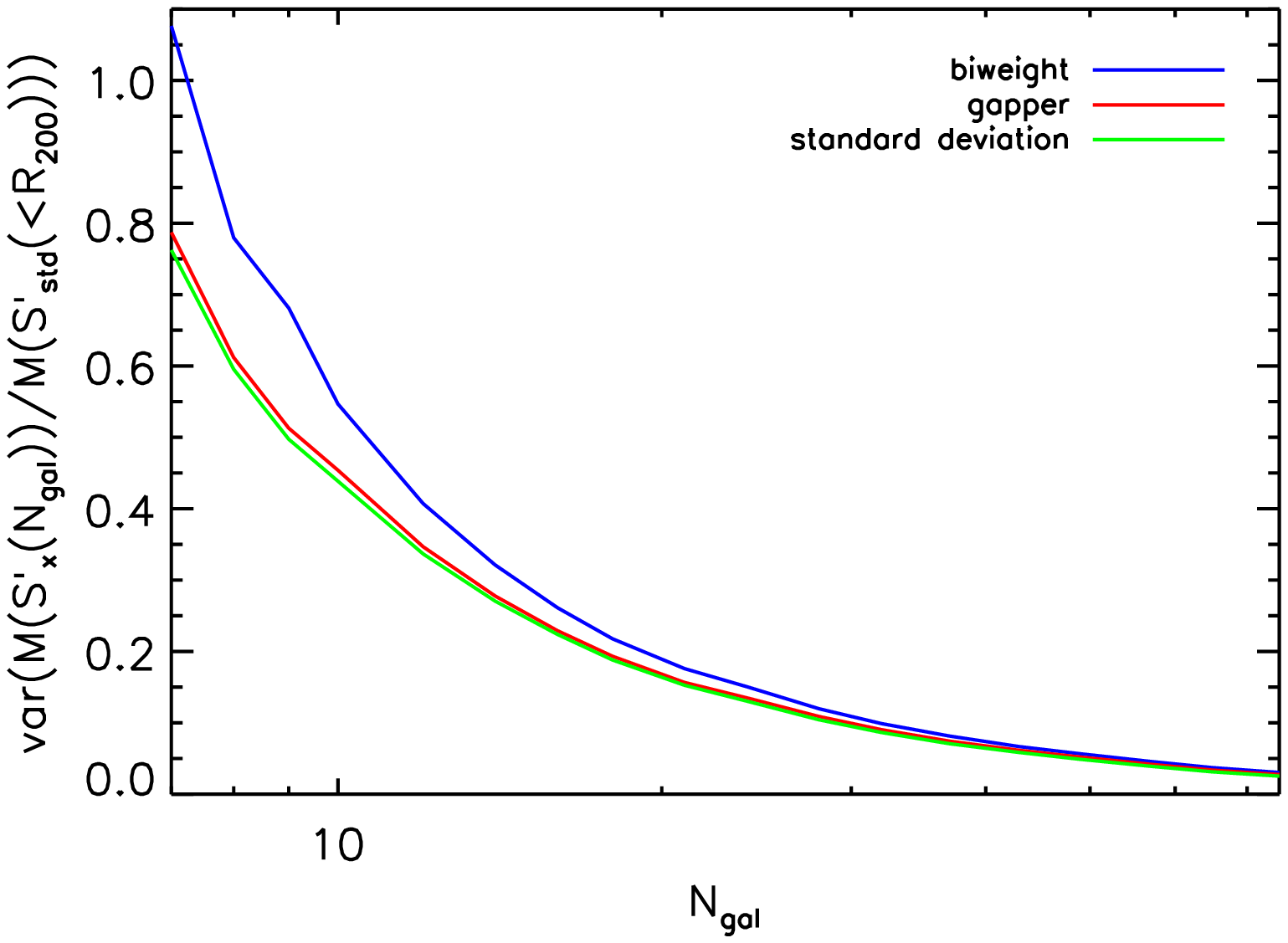}
\end{center}
\caption{Mean (left panels) and variance (right panels) of $M\left(\Sx(\ngal)
  \right)/M\left(\Sstd(<R_{200})\right)$ and $M\left(S'_x(\ngal)
  \right)/M\left(S'_{\rm std}(<R_{200})\right)$, which represent the standard
  mass estimator, eq.\ \ref{eq:fs_eq}, applied to normal and unbiased velocity
  dispersion estimators, standard deviation (green), gapper (red), and biweight
  (blue).  The theoretical expectation for
  $M\left(\Sstd(\ngal)\right)/M\left(\Sstd(<R_{200})\right)$ is represented with
  the black dashed line. }
  \label{fig:m_b_sim_f}
\end{figure*}

The mass of a GC is not a direct observable. When we estimate the cluster mass
using velocity dispersion estimates, we basically apply a function
$M$--$\sigma_{\rm 1D}$ that has been previously calibrated either in simulations
or using observations. Any non-linear transformation of $\sigma_{\rm 1D}$ will
introduce a bias similar to the one that we have discussed for the $S_{\rm X}$
estimators, which will be more significant in the low-$\ngal$ regime.

Following equation~\ref{eq:smg}, once we have obtained an estimate of the
velocity dispersion ($\Sx$), the mass of the cluster can be computed as
\begin{equation}
\frac{M(\Sx)}{10^{15}\, \mathrm{M}_{\odot}}=\left( \frac{\Sx}{A} \right)^{\frac{1}{\alpha}}.
\label{eq:fs_eq}
\end{equation}
with parameters $A=1177.0$\,km\,s$^{-1}$ and $\alpha=0.364$
\textbf{constrained by \cite{munari13} using the biweight as velocity
  dispersion estimator}. 

However, and in analogy to what we have seen in Sect.~\ref{sec:bias}, even if we
use an unbiased estimator for the velocity dispersion, and due to the fact that
equation~\ref{eq:fs_eq} contains a non-linear function of the variance, we
expect a statistical bias with some dependence at low $\ngal$. Using the results
from Appendix~\ref{apendix:ap1}, as the transformation to obtain the mass is of
the type $f(v) \propto v^{1/(2\alpha)}$ with $1/\alpha \sim 3$, we can predict
the amount of bias.

As for the study of the dispersion estimators, in the case of mass estimators we
also select as the reference velocity dispersion that one estimated with the
standard deviation using all the galaxies within $R_{200}$. In the top row of
Fig.~\ref{fig:m_b_sim_f} we show the results of the bias (left panel) and
associated variance (right panel) of the mass estimator based on
eq.~\ref{eq:fs_eq}, using $\Sx(\ngal)$ as the velocity estimator. This case is
noted as $\left< M(\Sx) / M(\Sstd(<R_{200}))\right>$, and $M(\Sstd(<R_{200}))$
represents the mass obtained using eq.~\ref{eq:fs_eq} for the input value of
$\Sstd(<R_{200})$.

We see that this mass estimator $M(\Sstd)$ is positively biased by a factor
\begin{equation}
  \frac{1-2\alpha}{ 4\alpha^2 (\ngal-1)},
\end{equation}
as predicted by equation~\ref{eq:ap5} in Appendix~\ref{apendix:ap1}. In a
similar way, the $M(S_{\rm bwt})$ (blue line) and $M(S_{\rm gap})$ (red line)
mass estimators are also biased positively biased.

For comparison, the bottom row of Fig.~\ref{fig:m_b_sim_f} shows the equivalent
results of the bias (left panel) and associated variance (right panel) of the
mass estimator but now using $\Sxprime(\ngal)$ as the input velocity
estimation. These quantities are represented as $\left< M(\Sxprime) /
M(\Sstd(<R_{200})) \right>$. As anticipated, even if we use an
unbiased velocity dispersion estimator, the non-linearity of the
mass--velocity dispersion relation results in a biased mass estimate.
It is interesting to note that the velocity dispersion bias is propagated into
the mass bias, as seen by comparing the top and bottom panels on that figure. On
one hand, as the $S_{\rm gap}$ is independent from $\ngal$, in the
transformation from the normal estimator to the bias-corrected one the only
thing that changes is the normalisation (i.e. a constant factor, and therefore,
the variance does not increase). On the other hand, the mass bias of $\Sstd$
(top panel) is mitigated by the fact that the normal standard deviation tends to
underestimate the velocity dispersion at low $\ngal$. This effect is not present
in the $M(S'_{\rm std})$ profile (bottom panel) because $S'_{\rm std}$ is, by
construction, unbiased.
Focusing our attention on the variance of these mass estimators, shown in the
right panels of Fig.~\ref{fig:m_b_sim_f}, we see that for $M(\Sx)$ (top panel),
the standard deviation has the lowest variance, whereas the gapper and biweight
show almost the same behaviour as functions of $\ngal$. Instead, in the bottom
panel the three $Var( M(\Sxprime) )$ functions show a behaviour similar to what
we see for the dispersion estimators. The gapper variance remains almost
untouched, and the standard deviation behaves like the gapper, whereas the
biweight has the higher variance.

As done in Sect.~\ref{sec:bias}, we have proposed a parametric description of
the bias as a function of $\ngal$, based on the analytic form of the bias for
the standard deviation case. We also use here three parameters ($E$,$F$ and
$\gamma$) in order to apply it to the gapper- and biweight-based mass
estimators:
\begin{equation}
\label{eq:fit_efg}
\frac{1 - E\alpha}{(E\alpha)^2 (\ngal-1)^\gamma} + F
\end{equation}
The best-fit parameters describing the bias for the unprimed $M(\Sx)$ and primed
$M(\Sxprime)$ mass estimators are listed in Tables~\ref{table:Fs_par_sim} and
\ref{table:Fs'_par_sim}, respectively.

Once we have fitted for this bias, we can propose bias-corrected mass estimators
for those two cases, by defining
\begin{align}
\label{eq:M_M'_eq}
M'\left( \Sx(\ngal) \right) = M\left(\Sx(\ngal) \right)\left[\frac{1-E\alpha}{(E\alpha)^2(N_{\rm gal}-1)^\gamma}+F\right]^{-1}\\
M'\left( \Sxprime(\ngal) \right) = M\left(\Sxprime(\ngal) \right)\left[\frac{1-E'\alpha}{(E'\alpha)^2(N_{\rm gal}-1)^{\gamma^\prime}}+F'\right]^{-1}. \label{eq:M_M"_eq}
\end{align}
Following the same convention for the notation adopted in previous sections,
hereafter these bias-corrected estimators are represented with a ``prime'',
i.e. $M'(\Sx(\ngal))$ and $M'(\Sxprime(\ngal))$.

\begin{table}
\caption{Best-fit parameters for the function describing the bias in $\left<
  M(\Sx)/M(\Sstd(<R_{200})) \right>$ for simulated clusters, as described in
  equation~\ref{eq:fit_efg}. }
\label{table:Fs_par_sim} 
\centering 
\begin{tabular}{c c c c} 
\hline\hline 
\noalign{\smallskip}
\, & $BWT$ & $GAP$ & $STD$ \\
\hline
\noalign{\smallskip}
$E$ & $2.36\pm0.06$     & $1.49\pm0.03$     & $1.97\pm0.07$\\ 
$F$ & $0.058\pm0.004$ & $0.023\pm0.002$ & $0.003\pm0.002$\\
$\gamma$ & $0.7\pm0.1$         & $1.17\pm0.04 $    & $1.15\pm0.09$ \\
\noalign{\smallskip}
\hline 
\end{tabular}
\end{table}
\begin{table}
\caption{Best-fit parameters for the function describing the bias in $\left<
  M(\Sxprime)/M(\Sstd(<R_{200})) \right>$ for simulated clusters, as described
  in equation~\ref{eq:fit_efg}. }
\label{table:Fs'_par_sim} 
\centering 
\begin{tabular}{c c c c} 
\hline\hline 
\noalign{\smallskip}
\, & $BWT$ & $GAP$ & $STD$ \\
\hline
\noalign{\smallskip}
$E'$ & $1.31\pm0.03$ & $1.50\pm0.03$    & $1.53\pm0.03$\\ 
$F'$ & $ 0 $                   & $ 0 $                      & $ 0 $\\
$\gamma^\prime$ & $1.24\pm0.03$ & $ 1.17\pm0.04 $  & $1.11\pm0.04 $ \\
\noalign{\smallskip}
\hline 
\end{tabular}
\end{table}

\begin{figure*}
\begin{center}%\input{figure2}
\includegraphics[width=0.49\textwidth]{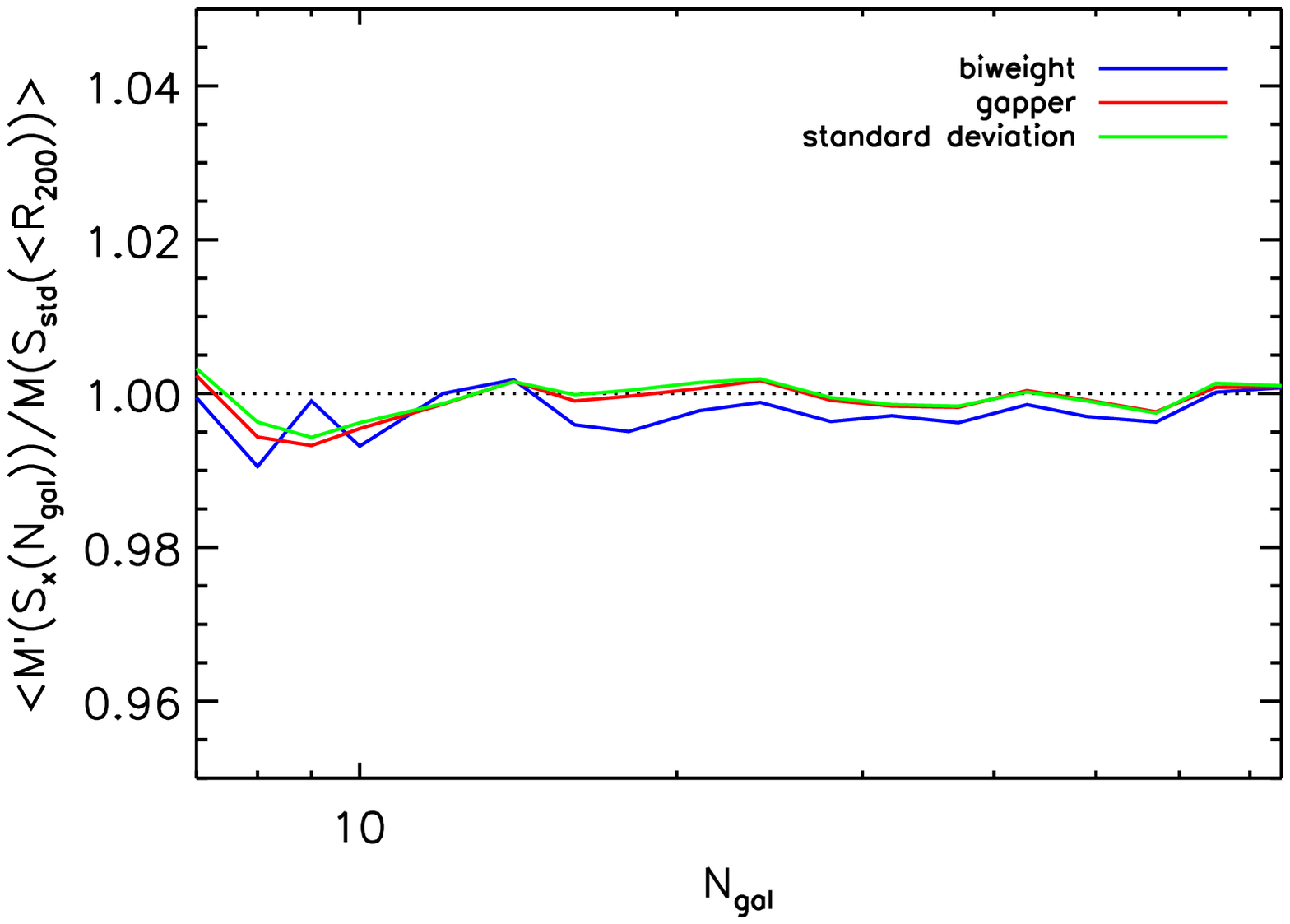}
\includegraphics[width=0.49\textwidth]{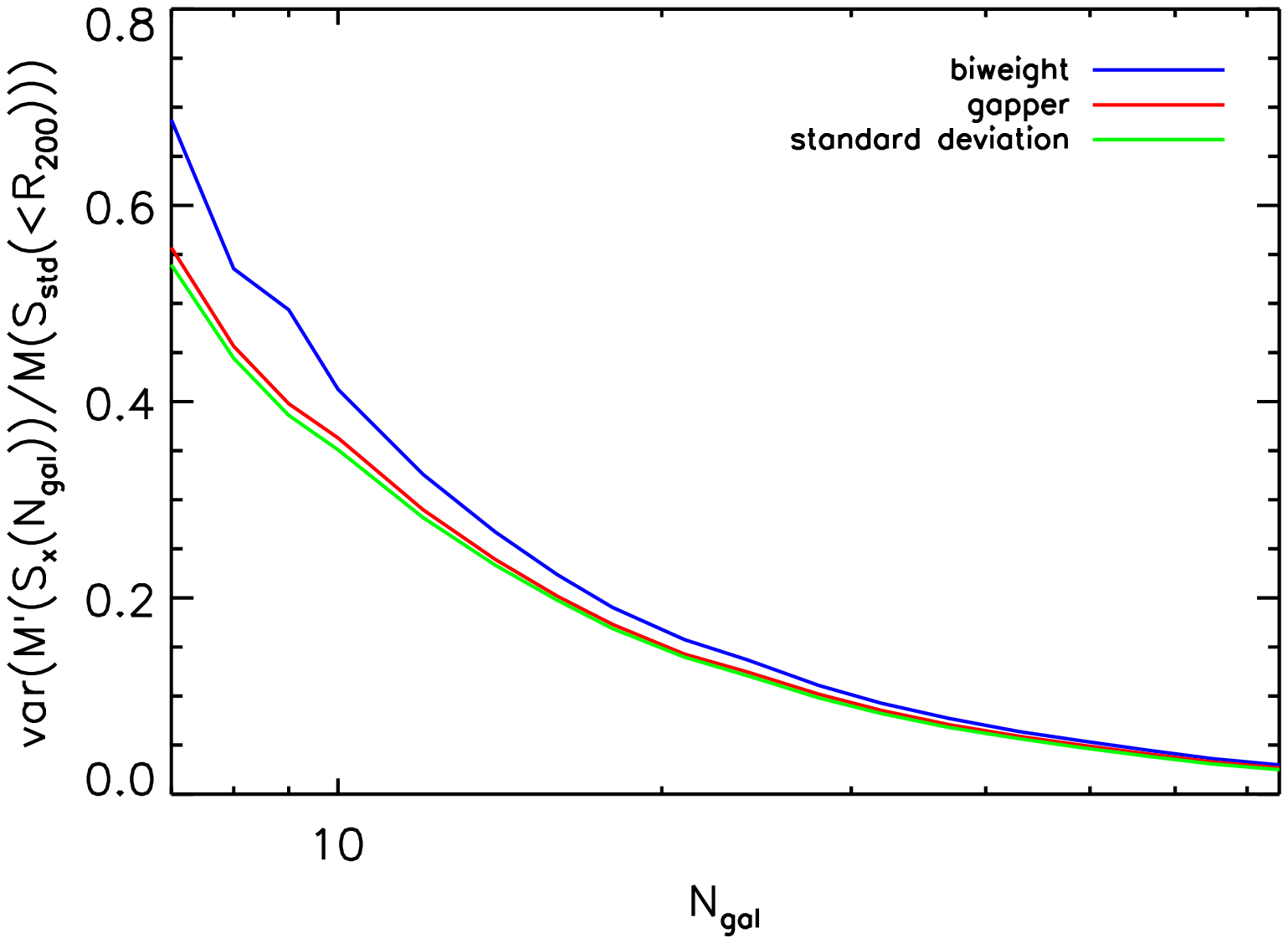}
\includegraphics[width=0.49\textwidth]{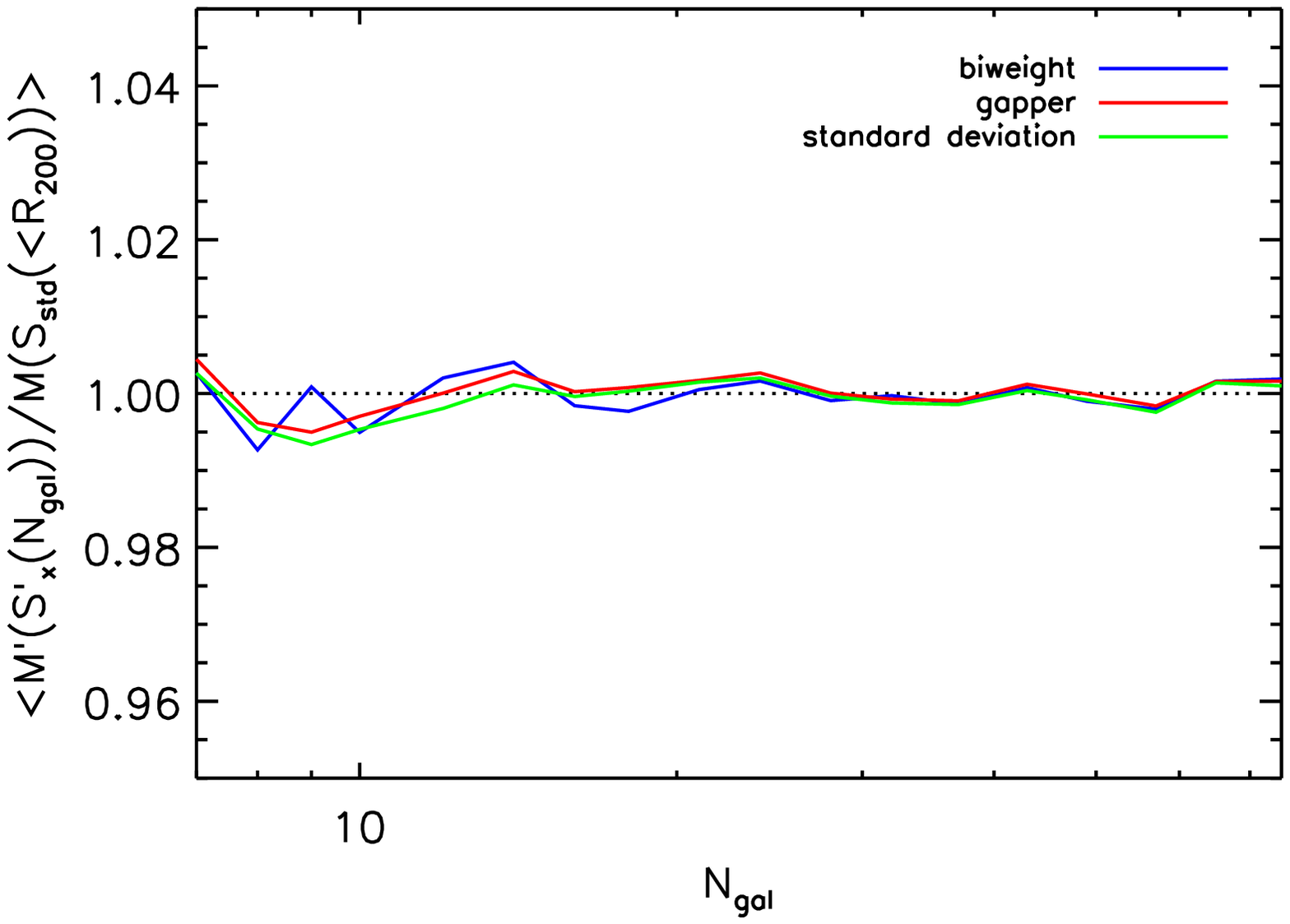}
\includegraphics[width=0.49\textwidth]{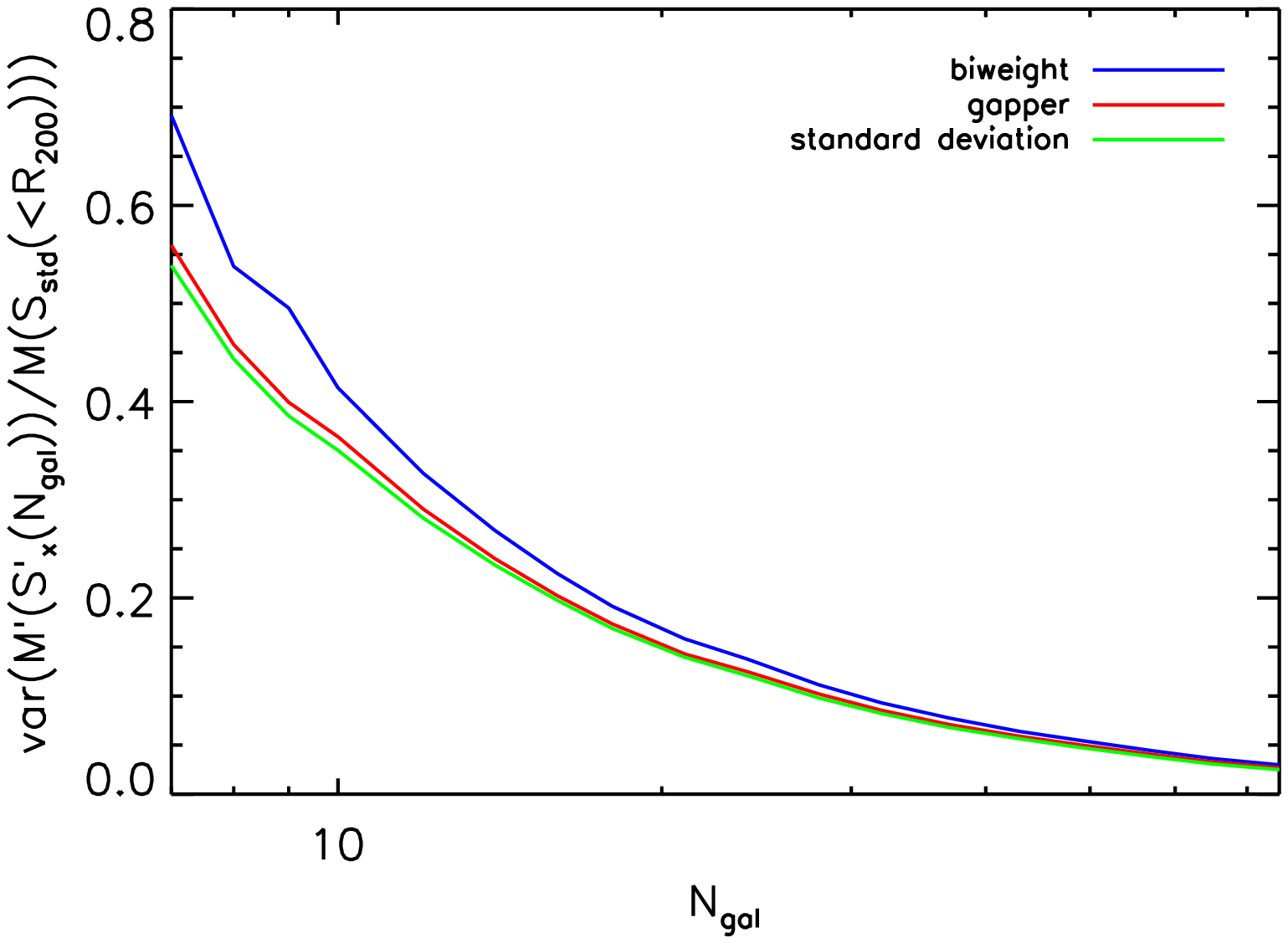}
\end{center}
\caption{Mean (left panels) and variance (right panels) of the unbiased mass
  estimators $M'(\Sx)/M'(\Sstd(<R_{200}))$ (top row) and
  $M'(\Sxprime)/M'(S'_{\rm std}(<R_{200}))$ (bottom row). In all panels, we show
  the results for the standard deviation (green), gapper (red), and biweight
  (blue) estimators. }
  \label{fig:m_b_sim_M}
\end{figure*}

Fig.~\ref{fig:m_b_sim_M} shows the bias and variance of $M'(\Sx)$ (top panel)
and $M'(\Sxprime)$ (bottom panel). Both estimators are actually unbiased by
construction. Concerning their variance, as expected, the biweight has the
largest variance, whereas standard deviation behave similarly to the gapper but
sill remaining as the lowest variance estimator. Analytical
  fits to the dependence of the variance as a
  function of $\ngal$ are given in Appendix~\ref{apendix:ap4}.

\subsection{Physical biases in the $M_{200}$ estimation}
\label{subsec:mass_bias}

In Sect.~\ref{sec:bias2} we explained how biases in velocity dispersion
estimation could appear by taking into account only the more massive cluster
members of the cluster, or by sampling a fraction of the virial radius
$R_{200}$. These biases due to the physics of galaxy clusters are also
propagated to the mass estimation.

\begin{figure*}
\begin{center}
\includegraphics[width=0.49\textwidth]{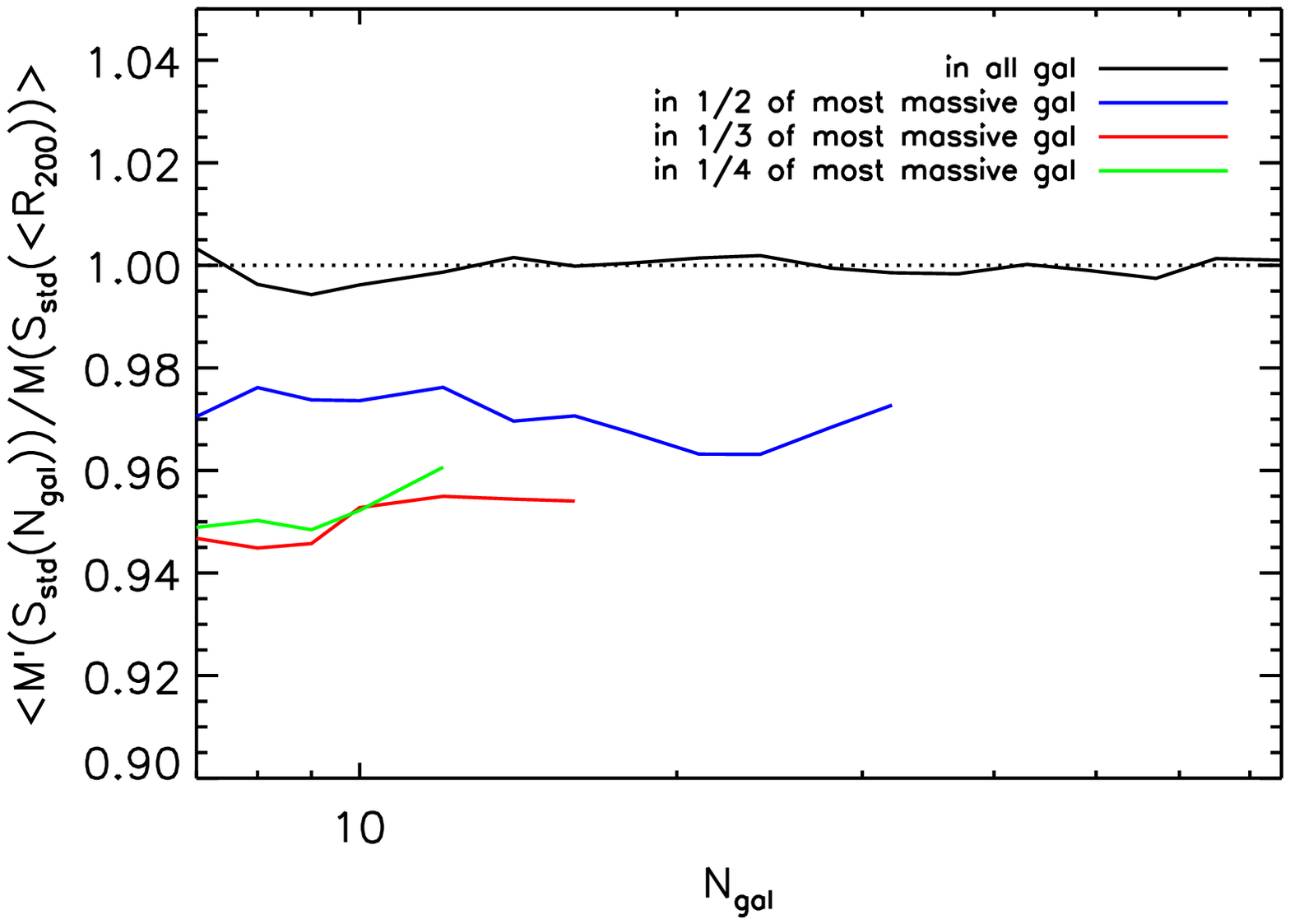}
\includegraphics[width=0.49\textwidth]{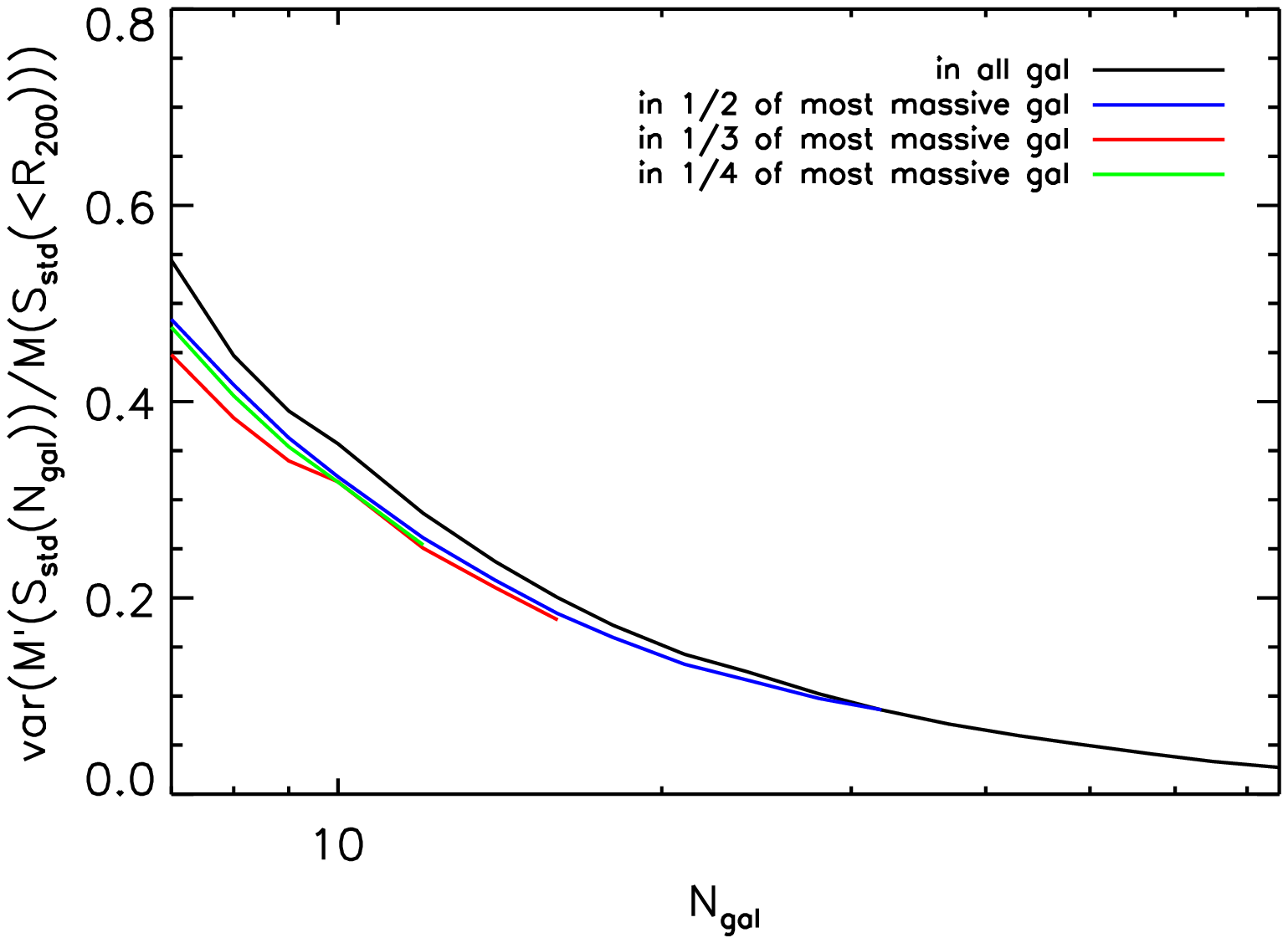}
\includegraphics[width=0.49\textwidth]{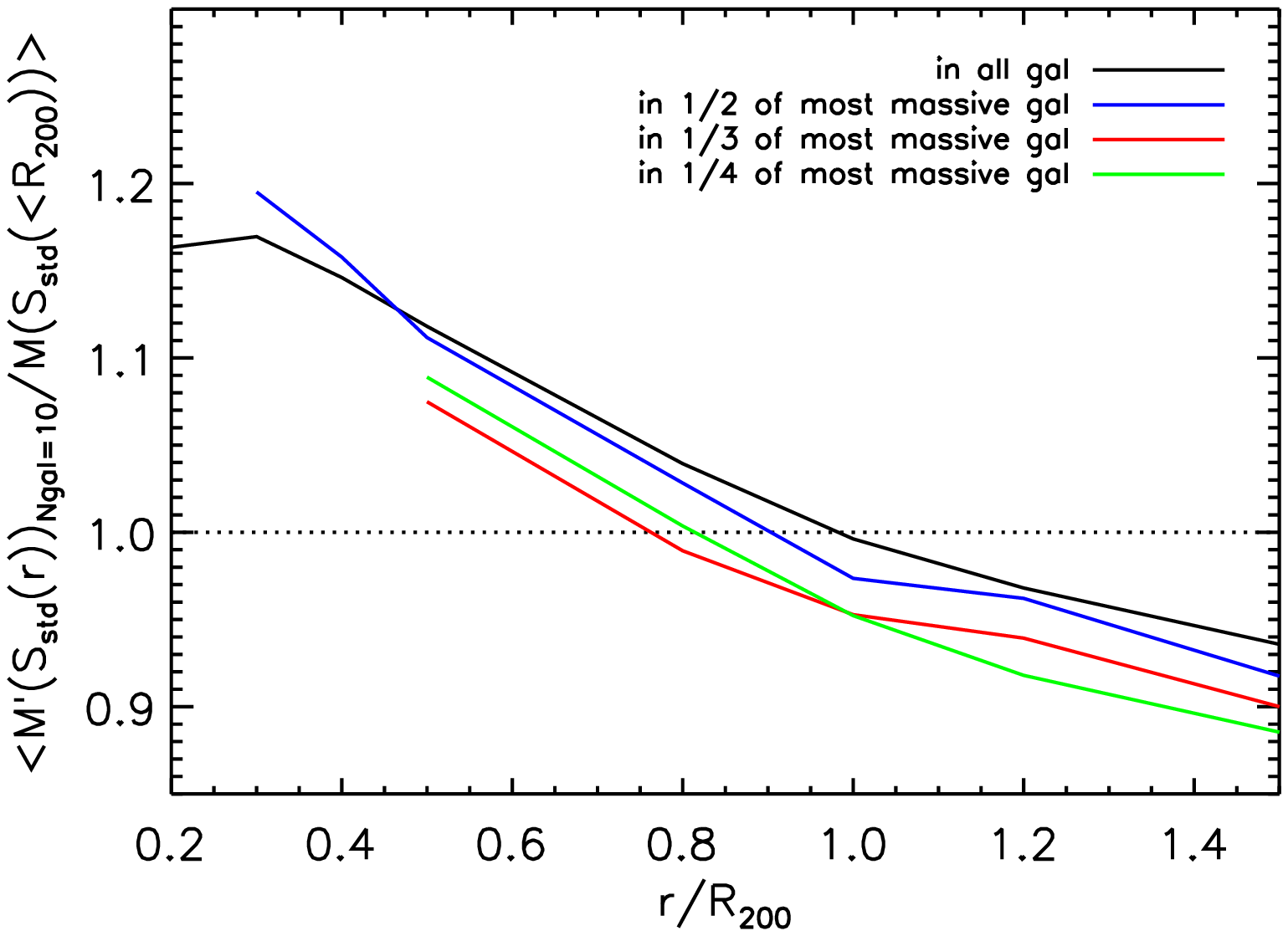}
\includegraphics[width=0.49\textwidth]{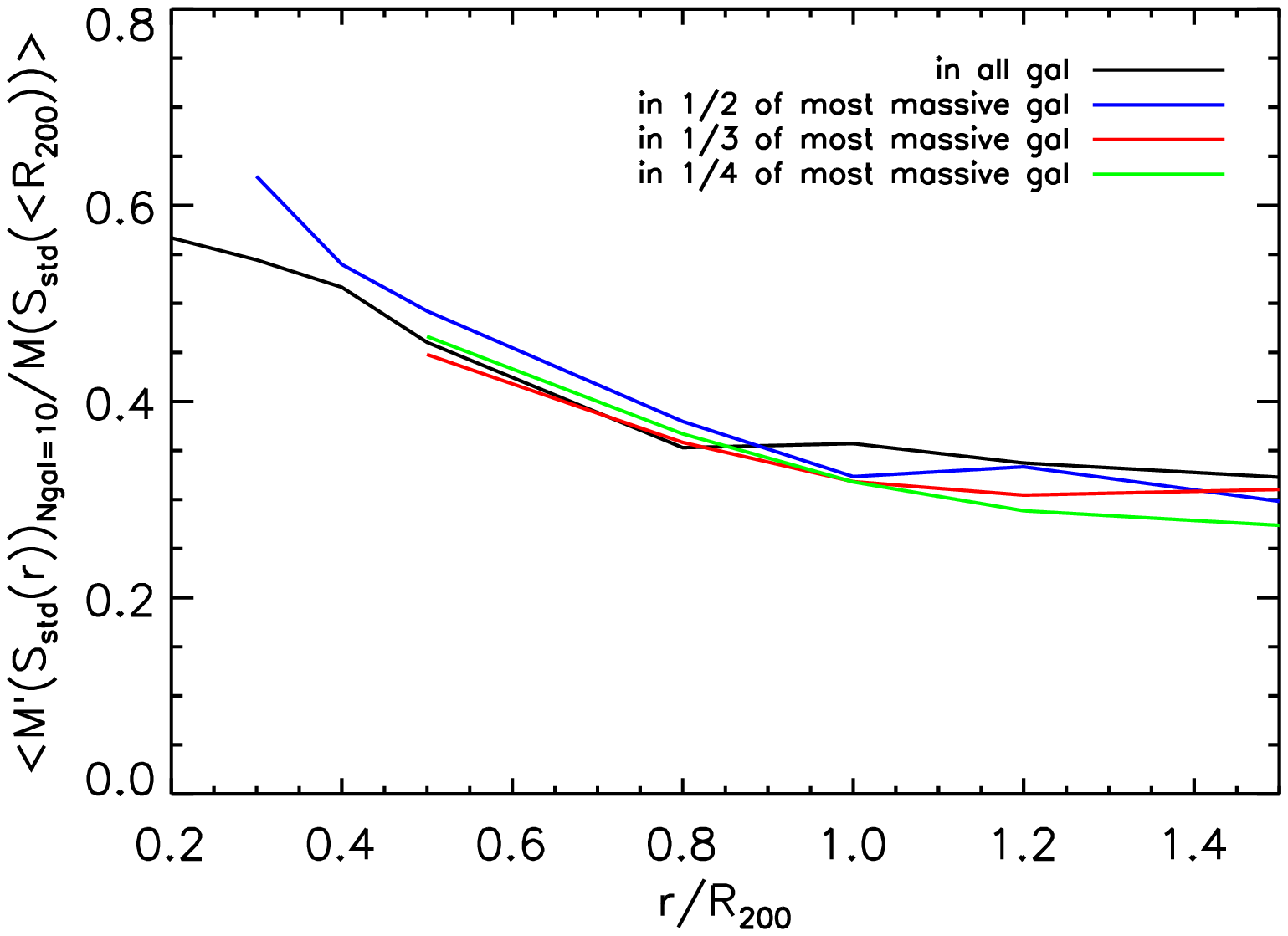}
\end{center}
\caption{Mean and variance of the $M'(\Sx)/M'(\Sstd(<R_{200}))$ mass estimator,
  as a function of the number of galaxies $\ngal$ (top panels), and as a
  function of the aperture radius $r$ (bottom panels). In black, blue, green,
  and red are represented the fraction that includes $100\%$, $1/2$, $1/3$ and
  $1/4$ of the most massive galaxies, respectively.}
  \label{fig:m_b_sim_physic}
\end{figure*}

In the top panels of Fig.~\ref{fig:m_b_sim_physic} we show how choosing galaxies
from the subset of the most massive ones also introduces a bias the mass
estimation. For illustration purposes, we show the effect on the $M'(\Sx)$
estimator only, but a similar figure can be generated for $M'(\Sxprime)$.
We find that the mass could be underestimated up to a $5\,\%$ when using $1/4$
of the sample containing the most massive galaxies. It is clear that the small
biases in the velocity estimation are now amplified, especially at low $\ngal$.

The aperture effect on the mass estimators is also shown in the bottom panel of
Fig.~\ref{fig:m_b_sim_physic}. Also in this case, all the biases are, as
expected, bigger than in the case of the velocity dispersion with a profile that
prevents steeper sampling going to bigger apertures. From the variance point of
view it increases in the core of the cluster and remains almost untouched for $r
\geq R_{200}$.

It is evident that, owing to the high variance of the mass estimation, the
combination of these effects may be considered negligible for a single cluster
mass determination, but in order to determine the mean bias of a scale relation
it is very important to obtain the most accurate mass estimation possible.

\section{Applying corrections to a realistic case}
%\label{subsec:sim_corrections}

\label{subsec:M_recipe}
\begin{figure*}
\begin{center}
\includegraphics[width=0.49\textwidth]{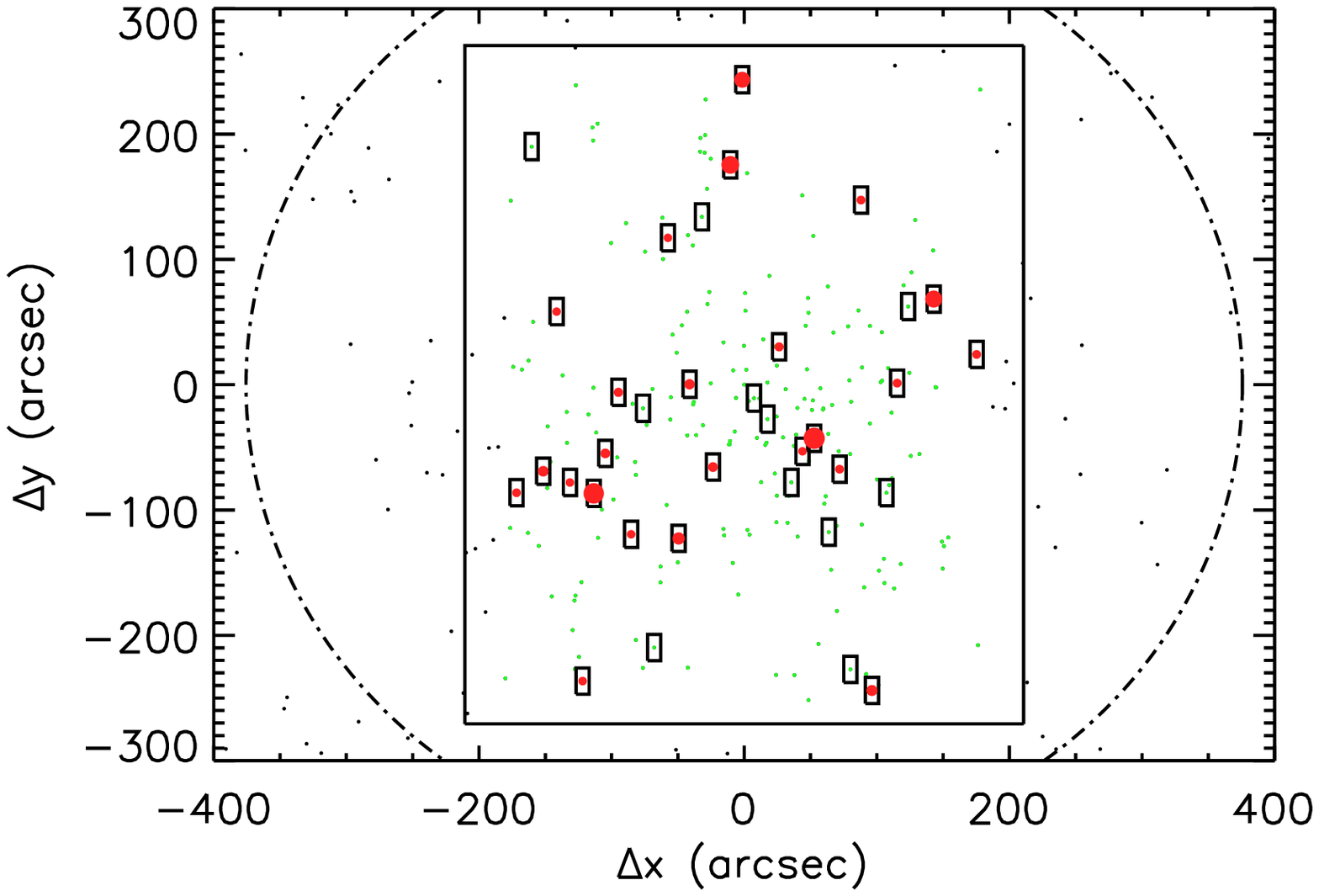}
\includegraphics[width=0.49\textwidth]{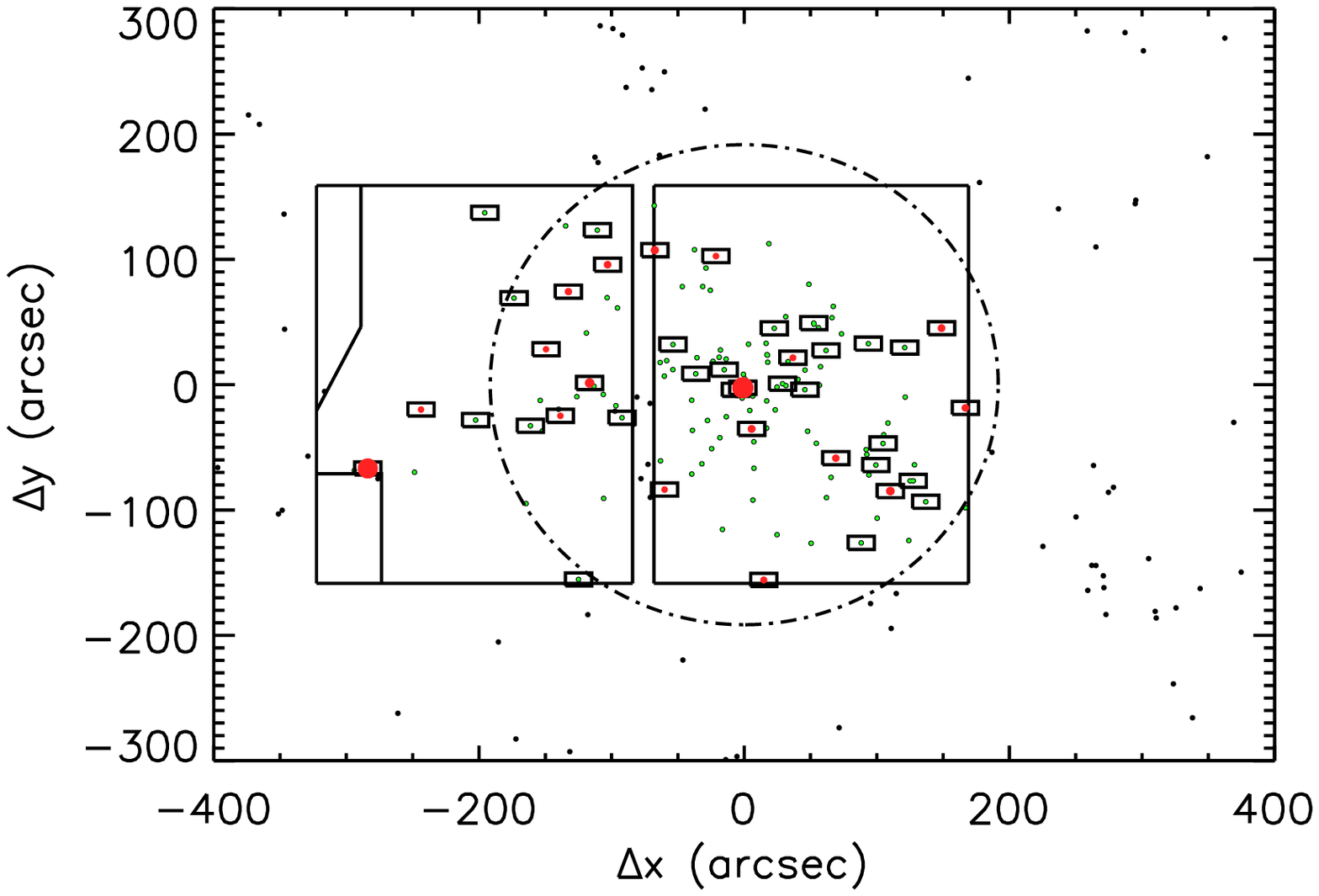}
\end{center}
\caption{Left panel: Example of DOLORES/TNG simulated mask. Right panel: Example
  of OSIRIS/GTC simulated mask. See text for details. Black and green dots
  represent cluster members outside or inside the telescope field of view,
  respectively. Black rectangles are the slitlets, and the red points those
  considered line-of-sight velocity measurements (the symbol size is
  proportional to the mass of each galaxy with respect to the most massive one
  in the mask). The circle in dotted line represents the projected $R_{200}$
  radius. }
  \label{fig:masks}
\end{figure*}

In this section, we show how well we can retrieve a bias-corrected velocity
dispersion and mass estimation for a simulated sample of galaxy clusters under
realistic observing conditions, but where only type 1
  interlopers are considered. We follow the methodology outlined in
Sect.~\ref{sec:recipe} and discussed in the previous sections. The basic steps
are:
\begin{enumerate}[label=\roman{*}., ref=(\roman{*})]
\item Use the unbiased estimator $\Sxprime$, defined in equation~\ref{eq:un_S},
  to estimate the cluster velocity dispersion.
\item Correct this velocity dispersion for the two physical effects (mass
  fraction and sampling aperture) described in the text, using
  Table~\ref{table:frac_bias} (or Fig.~\ref{fig:n_sigma_ahtqo}) and
  Table~\ref{tab:rad_bias} (or Fig.~\ref{fig:r_sigma_ahtqo}), respectively. The
  second correction requires a first-order estimation of $R_{200}$, which can be
  obtained from the $\Sxprime$ value from the previous step and the
  $\sigma_{200}$--$M_{200}$ relation in equation~\ref{eq:smg}.
\item Estimate the percentage of contaminants (interlopers) of the cluster
  members sample and correct, if needed, the velocity dispersion using the
  curves in Fig.~\ref{fig:interlopers2}. This provides the final velocity
  dispersion estimate, corrected for all effects described in the paper.
\item Compute the galaxy cluster mass by using the unbiased mass estimator
  $M'(\Sxprime)$ defined in equation~\ref{eq:M_M"_eq}, and using as input the
  corrected $\Sxprime$ value from the previous step.
\end{enumerate}

To perform this test, we decided to mimic the observational strategy that we
adopted in our Planck PSZ1 follow-up program carried out during a two-year
International Time Project (ITP13B/15A) \citep{nostro16,rafa18}. This
observational program has the aim of validating and characterising the unknown
Planck SZ sources of the PSZ1 catalog in the northern hemisphere. To do this
we used the DOLORES and OSIRIS spectrographs at the 3.5\,m Telescopio Nazionale
Galileo (TNG) and the 10.4\,m Gran Telescopio Canarias (GTC) respectively, both
located at Roque de los Muchachos Observatory (La Palma, Spain).

These facilities allow multi-object spectroscopy (MOS) observations that fit
very well with our aim. However, the high number of clusters to be observed
(about 200) and the need to obtain spectroscopy of very faint objects, $'r_{\rm
  mag}>22$, did not allow us to use more than one MOS mask or a couple of long
slits per cluster. For this reason we could only obtain a reduced number of
cluster members, $\ngal < 40$.  Our observational strategy started with the
photometric redshift estimation. Once the $z_{\rm phot}$ was determined, we
divided the GC sample into two redshift bins, $z\leq0.4$ and $z>0.4$, in order
to observe them at the TNG and the GTC respectively.  Here, we mimic the galaxy
selection procedure and the resultant galaxy catalogs. Owing to the different
fields of view (FOV) of the two instruments and the differences in the two-mask
designer software, we decided to implement both configurations following the
same prescription as described above. In Fig.~\ref{fig:masks} we show two
examples of a TNG and GTC mask scheme.  There are some differences in the two
configurations: i) TNG masks are always centered on the GC center while GTC
masks, owing to the gap between the two CCD, are shifted $\sim 100"$ to the
left; ii) the GTC mask designer tool is more precise than the TNG one. For this
reason the minimum distance between galaxies is $5.4"$ and $8"$ for GTC and TNG
masks respectively.

In order to simulate a cluster observation, we select each cluster at a random
orientation and fraction of visible galaxies.  Also, to mimic observational
issues such as spectral contamination or wrong mask centering, we set a random
number of slits as the effective catalog of measured radial velocities. We
repeated this procedure to obtain $100$ mock samples out of the $73$ GCs object
simulated in this study. For each of these samples we calculated the mean ratio
between the estimated and the reference cluster velocity dispersion. Because the
parameters of the scaling relation between $\sigma_{200}$--$M_{200}$,
eq.~\ref{eq:smg}, were constrained using the biweight estimation of the velocity
dispersion, $S_{\rm bwt}(<R_{200})$, we decided to use it as a reference
velocity dispersion for this analysis. For each of these samples we calculated
the mean ratio between the estimated and the reference velocity dispersion of
each cluster. Averaging over all the mock samples we obtained
\begin{align} \label{eq:S_pre_corr}
\begin{split}
\left< S_{\rm bwt}(\ngal,r)/S_{\rm bwt}(<R_{200}) \right> & = 0.96\pm0.03, \\
\left< S_{\rm gap}(\ngal,r)/S_{\rm bwt}(<R_{200}) \right> & = 0.99\pm0.03, \\
\left< \Sstd(\ngal,r)/S_{\rm bwt}(<R_{200}) \right>    & = 0.96\pm0.02.
\end{split}
\end{align}

Using the estimators $\Sxprime$ defined in eq.~\ref{eq:un_S} with the parameters
in Table~\ref{table:param_un_sim}, we corrected the bias due to the number of
galaxies. To correct the biases due to the GC physics, the first step is to
identify in which fraction of massive galaxies the detected cluster members
reside.
%\change{We decided to compare the mass of the less massive galaxy in the
%  effective catalogue with respect to the mean of the inferior limit of each
%  mass fraction of all $73$ simulated clusters}.
The second step is to calculate the aperture radius, which is the sampling
radius.  To do this, first we needed to estimate $R_{200}$. Based on a first
estimate of the cluster mass using $\Sx(\ngal)$, we can derive a first-order
approximation to that radius, noted as $R^{\rm X}_{200}$. This value is used to
apply the aperture correction shown on the right panel of
Fig.~\ref{fig:r_sigma_ahtqo} to $\Sxprime$. In our case, each velocity
dispersion was corrected individually after averaging over all the clusters, and
then averaging over all the $100$ configurations to obtain
\begin{align} \label{eq: S_post_corr}
\begin{split}
\left< S'_{\rm bwt}(N_{\rm gal},r)/S_{\rm bwt}(<R_{200})\right> & = 1.00\pm0.03, \\
\left< S'_{\rm gap}(N_{\rm gal},r)/S_{\rm bwt}(<R_{200})\right> & = 1.00\pm0.02, \\
\left< S'_{\rm std}(N_{\rm gal},r)/S_{\rm bwt}(<R_{200})\right> & = 1.00\pm0.02,
\end{split}
\end{align}
which represent bias-corrected estimates of the velocity
dispersion. We note that these quantities are referred to the
  velocity dispersion computed in the cylinder $S_{\rm
    bwt}(<R_{200})$, and thus they include type 1 interlopers as
  described in Sect.~\ref{subsec:bias_inter_ref}. If we want to
  correct those values from this effect, we have to multiply those
  values by the corrections factors quoted in that subsection, i.e.,
  $0.990$, $0.981$ and $0.985$ for the biweight, gapper and standard
  deviation estimators, respectively.

Using these velocity dispersion, we can now calculate the cluster masses,
$M(\Sxprime)$, obtaining
\begin{align} \label{eq:Fsp_post_corr}
\begin{split}
\left< M\left( S'_{\rm bwt}(N_{\rm gal},r)\right)/M\left(S_{\rm bwt}(<R_{200})\right)\right> & = 1.17\pm0.09, \\
\left< M\left( S'_{\rm gap}(N_{\rm gal},r)\right)/M\left(S_{\rm bwt}(<R_{200})\right)\right> & = 1.14\pm0.08, \\
\left< M\left( S'_{\rm std}(N_{\rm gal},r)\right)/M\left(S_{\rm bwt}(<R_{200})\right)\right> & = 1.13\pm0.07.
\end{split}
\end{align}
As shown above, these masses are overestimated, and in order to correct for this
bias, we have to use the primed mass estimator, $M'$, as explained in
Sec.~\ref{sec:s_mass_bias}. In this case, we obtain
\begin{align} \label{eq:Msp_pre_corr}
\begin{split}
\left< M'\left( S'_{\rm bwt}(N_{\rm gal},r)\right)/M'\left(S_{\rm bwt}(<R_{200})\right)\right> & = 1.00\pm0.07, \\
\left< M'\left( S'_{\rm gap}(N_{\rm gal},r)\right)/M'\left(S_{\rm bwt}(<R_{200})\right)\right> & = 1.00\pm0.07, \\
\left< M'\left( S'_{\rm std}(N_{\rm gal},r)\right)/M'\left(S_{\rm bwt}(<R_{200})\right)\right> & = 1.00\pm0.06.
\end{split}
\end{align}

It is also interesting to compare these mass estimates with the true mass,
$M_{200}$ directly estimated from the simulation as the mass of all particles
within $R_{200}$. Those values were used to constrain the parameters in
equation~\ref{eq:smg} \citep{munari13}. The direct estimation of the velocity
dispersion using $\Sx$ leads to a biased estimation of $M_{200}$:
\begin{align} \label{eq:FsMT_pre_corr}
\begin{split}
\left< M\left( S_{\rm bwt}\,(N_{\rm gal},r)\right)/M_{200}\right> & = 1.07 \pm 0.09, \\  
\left< M\left( S_{\rm gap}\,(N_{\rm gal},r)\right)/M_{200}\right> & = 1.13 \pm 0.08, \\ 
\left< M\left( S_{\rm std}\,(N_{\rm gal},r)\right)/M_{200}\right> & = 1.03 \pm 0.08,     
\end{split}
\end{align}

If the dispersion estimator is the corrected one, $\Sxprime$, we now retrieve a
mass that is almost unbiased:
\begin{align} \label{eq:FspMt_post_corr}
\begin{split}
\left< M\left( S'_{\rm bwt}\,(N_{\rm gal},r)\right)/M_{200}\right> & = 1.18 \pm 0.10, \\    
\left< M\left( S'_{\rm gap}\,(N_{\rm gal},r)\right)/M_{200}\right> & =  1.13 \pm 0.09, \\  
\left< M\left( S'_{\rm std}\,(N_{\rm gal},r)\right)/M_{200}\right> & =  1.13 \pm 0.08,     
\end{split}
\end{align}

Finally, $M'$ provides the final, unbiased estimate of $M_{200}$:
\begin{align} \label{eq:MspMT_post_corr}
\begin{split}
\left< M'\left( S'_{\rm bwt}\,(N_{\rm gal},r)\right)/M_{200}\right> & =  1.00 \pm 0.08, \\  
\left< M'\left( S'_{\rm gap}\,(N_{\rm gal},r)\right)/M_{200}\right> & =  1.00 \pm 0.07, \\  
\left< M'\left( S'_{\rm std}\,(N_{\rm gal},r)\right)/M_{200}\right> & =  0.99 \pm 0.07, \\  
\end{split}
\end{align}
Again, for this example these quantities are referred to a ``true''
  mass computed from the velocity dispersion computed in the cylinder $S_{\rm
    bwt}(<R_{200})$. If we want to isolate the effect of type 1 interlopers, then the true $M_{200}$ has to be computed using the velocity
  dispersion in a sphere of radius $R_{200}$, and the primed velocity
  dispersion estimates will have to be corrected also for the same
  bias.

These numbers show that also in this realistic situation, the proposed set of
corrected estimators are able to recover an unbiased estimate for both the
velocity dispersion and mass of the galaxy clusters.

\section{Conclusions}
\label{subsec:end}
In this article, we have used 73 simulated GCs from hydrodynamic simulations
with AGN feedback and star formation. We have taken into account three different
estimators: the biweight, the gapper, and the standard deviation. We have
focused on the limit of a low number of galaxy members ($\ngal < 75$), and
studied the bias and error (variance) of each estimator.

This study presents a detailed study of three techniques to estimate velocity
dispersion quantifying possible biases brought about by their definition and
observational limits. In a future paper, we shall apply the optimal technique
with the corresponding corrections to a real sample of GCs in order to estimate
bias-corrected dynamical masses and compare them with those evaluated using
different proxies.

We propose a recipe with the aim of estimate reliable velocity dispersion and
mass estimators with the lowest bias and variance as possible in the low-$\ngal$
regime. We constructed unbiased estimators based on the standard deviation,
biweight, and gapper while correcting for their $\ngal$ dependence,
$\Sxprime$. In this case, we focused our attention on the variance of these
estimators. Although asymptotically the three estimators have the same variance,
in the range of a number of galaxies in which we are interested, $\ngal<40$, the
corrected biweight has an even higher variance with respect to the normal
biweight and, consequently, with respect to the other two estimators. After the
bias correction, we see that the variance of the standard deviation and gapper
are compatible for $\ngal \geq 20$, whereas for $\ngal < 20$ the corrected
standard deviation is the lowest variance estimator.

We have also tested the robustness of the three $\Sxprime$ estimators when the galaxy
sample is contaminated by interlopers, considering both
gravitationally bound interlopers (type 1) and background/foregrounds galaxies (type 2).  
For type 1 interlopers, the bias in the velocity dispersion estimator
is found to be $\sim 2$\,\%, comparable to the other statistical biases
discussed above, and consistent with the results obtained by
other authors which include also type 2 interlopers in their analysis \citep[e.g.][]{mamon10}. This bias can be corrected
using a $2.7\sigma$ clipping technique \citep{YahilVidal1977}. For type 2 interlopers, here we explored the conservative
approach of assigning them a uniform velocity distribution, which is completely different from the true distribution of cluster members. 
Our results show that the three estimators are similarly affected, and that at first
order, the bias in $\Sxprime$ is roughly proportional to the percentage of
type 2 interlopers in our approach. Although comparable to the other effects discussed in
this paper, the contribution of type 2 interlopers could provide the main bias in the velocity dispersion estimation, specially for radii 
beyond $R_{200}$, if their fraction is as large as 10\,\% \citep[e.g.][]{saro13}. Finally, this fraction of
type 2 contaminants depends on the particular GCs member selection procedure and,
hence, can vary from survey to survey. Thus, a general formula based
only on $\ngal$ can not be provided.

We also studied how observational limitations influence the estimation of velocity
dispersion in GCs. We recognised the most likely sources of bias in i) the
selection effect due to the luminosity of GC members observed and hence in the
fraction of massive galaxies used to estimate the velocity dispersion; ii) the
aperture radius of the observation, and hence the fraction of the viral radius
explored. We saw that the bias increased for the smaller fraction of massive
galaxies. This bias was estimated to be around the $2\,\%$ when considering only
$1/4$ of the more massive galaxies.

Regarding the effect produced by sampling aperture, we found that the maximum
deviation was produced for an aperture radius of $0.3$--$0.4\, R_{200}$, which
is in agreement with \cite{sifon16} results.

We also tested the mass estimators defined by eq.~\ref{eq:smg} with the
parameters of \cite{munari13}. In this case, we observed that all three
estimators had a dependence on the number of galaxies, overestimating the
reference mass at low $\ngal$. The standard deviation profile can be
analytically derived by taking into account the fact that the mass is a
nonlinear function of the variance, as explained in
Appendix~\ref{apendix:ap1}. Also, in this case, we defined a new set of mass
estimators, $M(\Sx)$ and $M'(\Sxprime)$, correcting the mass estimator applied
to the biweight, gapper, and standard deviation, or to their corrected
counterparts, respectively. We constructed these estimators to be unbiased over
the entire range of $\ngal$ and found that the mass estimator based on the
standard deviation, whether corrected or not, was the lowest variance one.

The ultimate aim of this paper is not only to correct the velocity dispersion
and mass estimates of a single cluster, but also to develop a method to obtain
bias-corrected mean mass estimates in order to constrain the parameter of mass
scaling relations. For this reason, we also applied these techniques to a set of
mock observations to retrieve bias-corrected mean velocity dispersion and
masses. These types of analyses will be of relevance for
  precision cosmology analyses with large samples of galaxy clusters, in the
  light of forthcoming results from space missions as eROSITA or
  EUCLID.

%ACKNOWLEDGEMENTS
\begin{acknowledgements}
We thank Andrea Biviano for useful discussions and for the comments on the
draft. We also thank the anonymous referee for useful suggestions
  related to the discussion of interlopers. We are greatly indebted to Volker Springel for providing us with the
non-public version of GADGET-3. Simulations have been carried out in CINECA
(Bologna), with CPU time allocated through the Italian SuperComputing Resource
Allocation (ISCRA) and through an agreement between CINECA and the University of
Trieste. AF, RB, JB and JARM acknowledge financial support from the Spanish Ministry
of Economy and Competitiveness (MINECO) under the projects ESP2013-48362-C2-1-P,
AYA2014-60438-P and AYA2017-84185-P. 
\end{acknowledgements}

\bibliographystyle{aa} 
\bibliography{bibliography_nightmare.bib} 
\begin{appendix}
\section{Bias in the estimation of the nonlinear function of variance $g(v)$}
\label{apendix:ap1}
If we assume a random variable $x$ with probability $p_i$, we can estimate any
ordinary moment as:
\begin{equation}
E[x^j]=\sum_{i=1}^Nx_i^j p_i.
\label{eq:ap1}
\end{equation}
The moment of second order is the variance, $E[Var(x)]=E[v]=<x^2>_n-<x>^2_n$,
which is unbiased for any number of data, $N$.

In this appendix we explain how a bias is induced when a quantity is calculated
as a nonlinear function of variance, $m=g(v)$. We evaluate $E[g(v)]$, which is
the estimate of the function $g(v)$. We can use as the estimator of $m$ the
function $g$ applied at the estimated value of $E[v]$,
\begin{equation}
E[g(v)]=g(E[v]).
\label{eq:ap2}
\end{equation}
The estimator $g(E[v])$ will be unbiased only if the function $g(v)$ is
linear. In case that $m$ is a nonlinear function of the variance, e.g. standard
deviation and mass, we can write the mean of $g(E[v])$ as:
\begin{equation}
<g(E[v])>\simeq g(v) +g^\prime(v) <\Delta E[v]> +\frac{1}{2}g^{\prime\prime}(v) <(\Delta E(v))^2>.
\label{eq:ap3}
\end{equation}
If the variable $v$ is unbiased, we can assume that $\left<\Delta E[v]\right>=0$ and
\begin{equation}
\left<(\Delta E[v])^2\right>=var(E[v])=\frac{2v^2}{N},
\label{eq:ap4}
\end{equation}
which implies that the unbiased estimate of $g(v)$ is

\begin{equation}
E[g(v)]= g(v) -\frac{1}{2}g^{\prime\prime}(v) \frac{2v^2}{N}.
\label{eq:ap5}
\end{equation}
In the case of the standard deviation $\sigma = v^{1/2}$ we have
\begin{equation}
E[\sigma]=\sigma+\frac{1}{4}\frac{\sigma}{n},
\label{eq:ap6}
\end{equation}
which is the unbiased standard deviation estimator.

\section{Statistical bias and variance for velocity dispersion and
  mass estimators in the limit of a Gaussian distributions}
\label{apendix:ap2}
\begin{figure*}
\begin{center}
\includegraphics[width=0.49\textwidth]{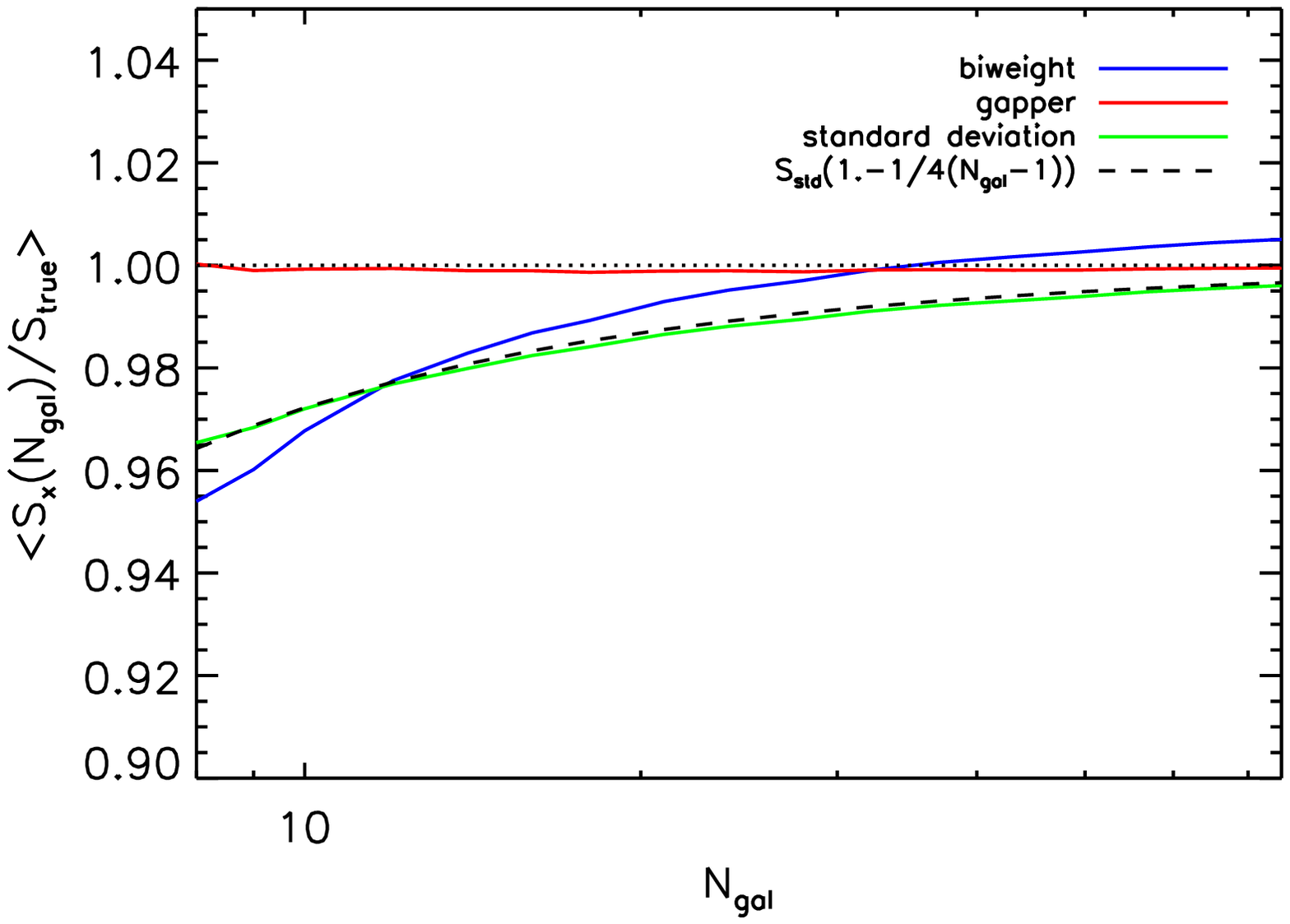}
\includegraphics[width=0.49\textwidth]{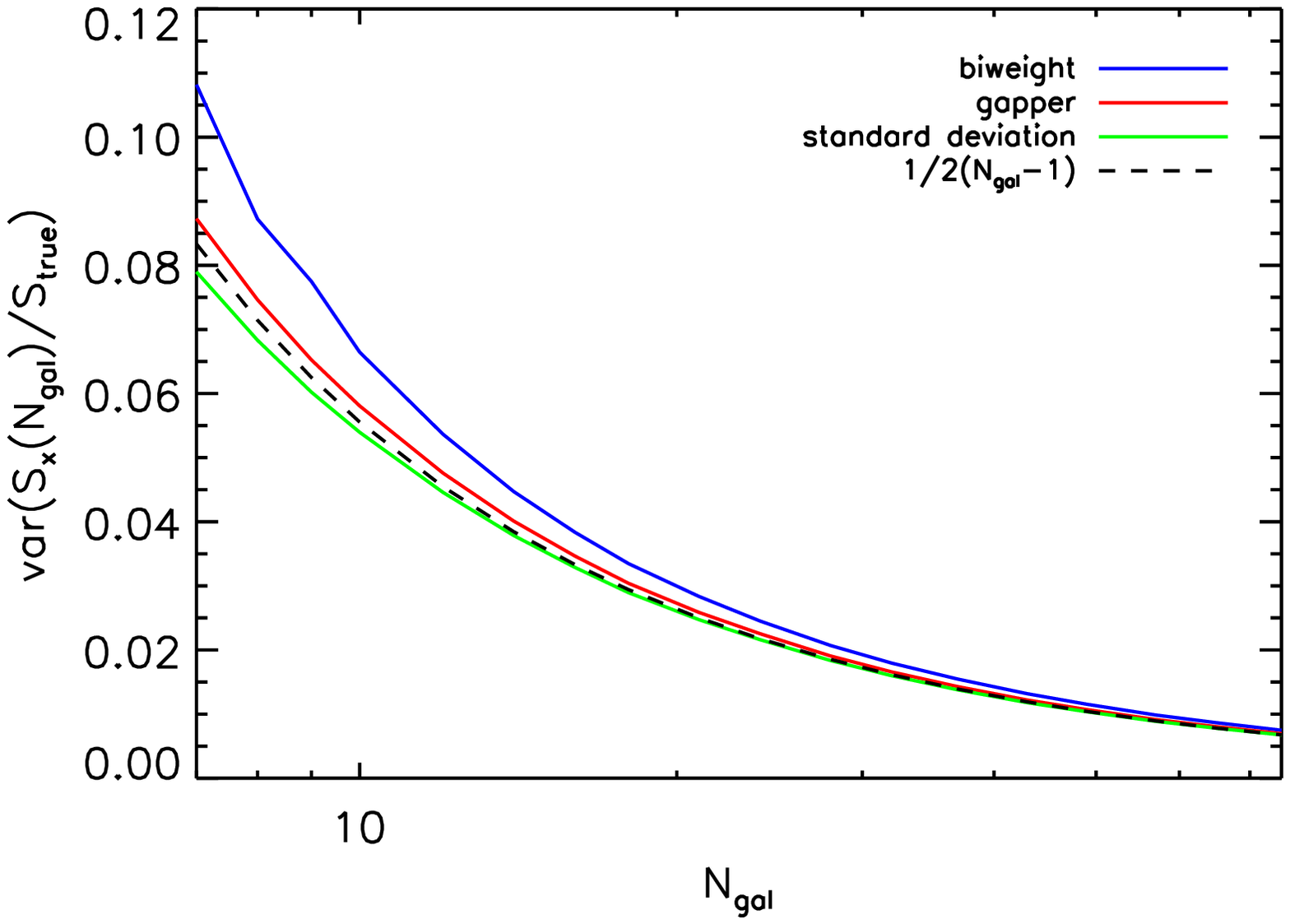}
\end{center}
\caption{Mean (left panel) and variance (right panel) of $\Sx(\ngal)$, as a function of the number of
  galaxies $\ngal$, for the case of Gaussian-distributed random
  velocities. Green, red, and blue lines represent respectively the
  behaviour of the standard 
  deviation, gapper, and biweight estimators.  }
  \label{fig:gauss_1}
\end{figure*}

In analogy to the cases discussed in the main text, in this appendix
we show how the velocity dispersion and mass estimators presented
above behave in the limit of a perfectly Gaussian velocity
distribution. To this end, we generate 73 Gaussian distributions
fixing the mean ($\mu = 0$), and the
dispersion $\sigma = \Strue$,  being $\Strue$ the dispersion
corresponding to $M_{200}$ though eq.~\ref{eq:smg} for each cluster in
this new simulated sample.

As in the main text, we explore 20 different values for $\ngal$,
between $\ngal=7$ and $\ngal=75$. We used $50\,000$ different
configuration for each $\ngal$ and each ``cluster'', thus using  $73 \times
50\,000$ velocity dispersion estimates normalised with respect to
$\Strue$. We note that by fixing the width of the Gaussian distribution, we
avoid the need to select a reference estimator and, hence, we can
investigate the absolute bias of each estimator with respect to the
same $\Strue$ value.

\subsection{Velocity dispersion estimators}
Fig.~\ref{fig:gauss_1} illustrates the behaviour of the (uncorrected) velocity estimators. We note that the gapper and
the standard deviation recover the true velocity dispersion,
asymptotically, for high $\ngal$. However, this is not the case for the biweight estimator,
which presents a small asymptotic bias. 
In the small number of galaxies limit, the
behaviour of the three estimators is very similar to the one
showed in Fig.~\ref{fig:n_sigma}. The standard deviation (green line)
follows, as expected, the exact analytic profile derived in
Appendix~\ref{apendix:ap1}, whereas the gapper (red
line) shows to be an almost unbiased estimator for any value of $\ngal$.
Table~\ref{table:param_un} presents the best-fit values of the parameters $D$,
$\beta$ and $B$ in equation~\ref {eq:parfit_sigma}, obtained now for
this case of Gaussian velocity distributions.

\begin{table}
\caption{Best-fit parameters for the functions describing $\Sx(\ngal)$
  in the case of Gaussian velocity distributions. }
\label{table:param_un} 
\centering 
\begin{tabular}{c c c c} 
\hline\hline 
\noalign{\smallskip}
\, & $BWT$ & $GAP$ & $STD$ \\
\hline
\noalign{\smallskip}
$D$ & $0.72\pm0.03$ & $0$ & $0.25$\\ 
$B$ & $-0.007\pm0.001$ & $0.0007\pm0.0002$ & $0$\\
$\beta$ & $1.28\pm0.03$ & $1$ & $1$\\
\noalign{\smallskip}
\hline %inserts single line
\end{tabular}
\end{table}

Concerning the variance, from a theoretical point of view we
would expect the standard deviation to be the lowest variance
estimator for a Gaussian distribution. This is confirmed in
Fig.~\ref{fig:gauss_1}, which shows that as expected, the standard
deviation follows the theoretical prescription given by 
\begin{equation}
var(\Sstd(\ngal))=\frac{S_{\rm true}^2}{2(\ngal -1)}.
\label{eq:v_sig_teoric}
\end{equation}

\begin{figure*}
\begin{center}
\includegraphics[width=0.49\textwidth]{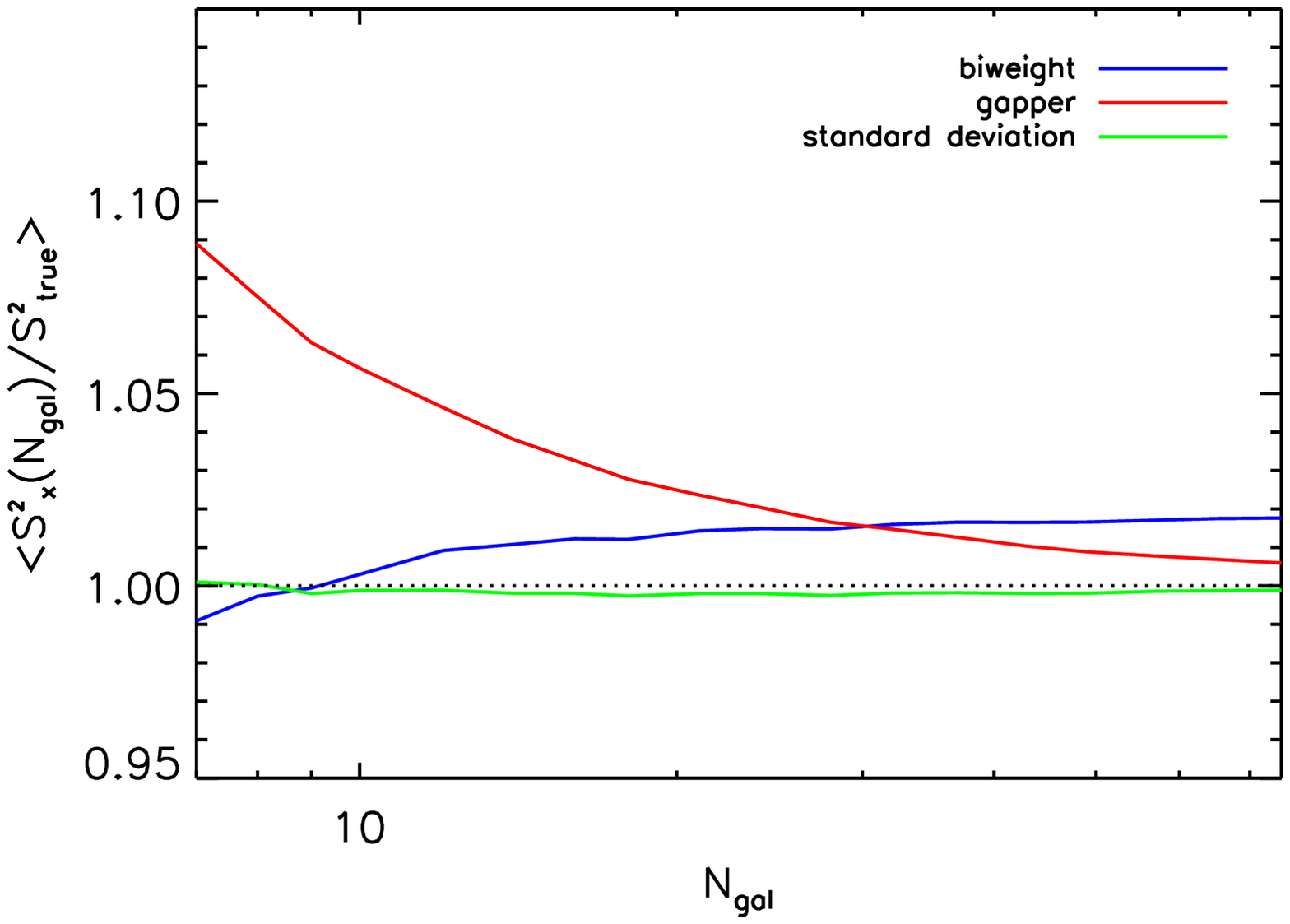}
\includegraphics[width=0.49\textwidth]{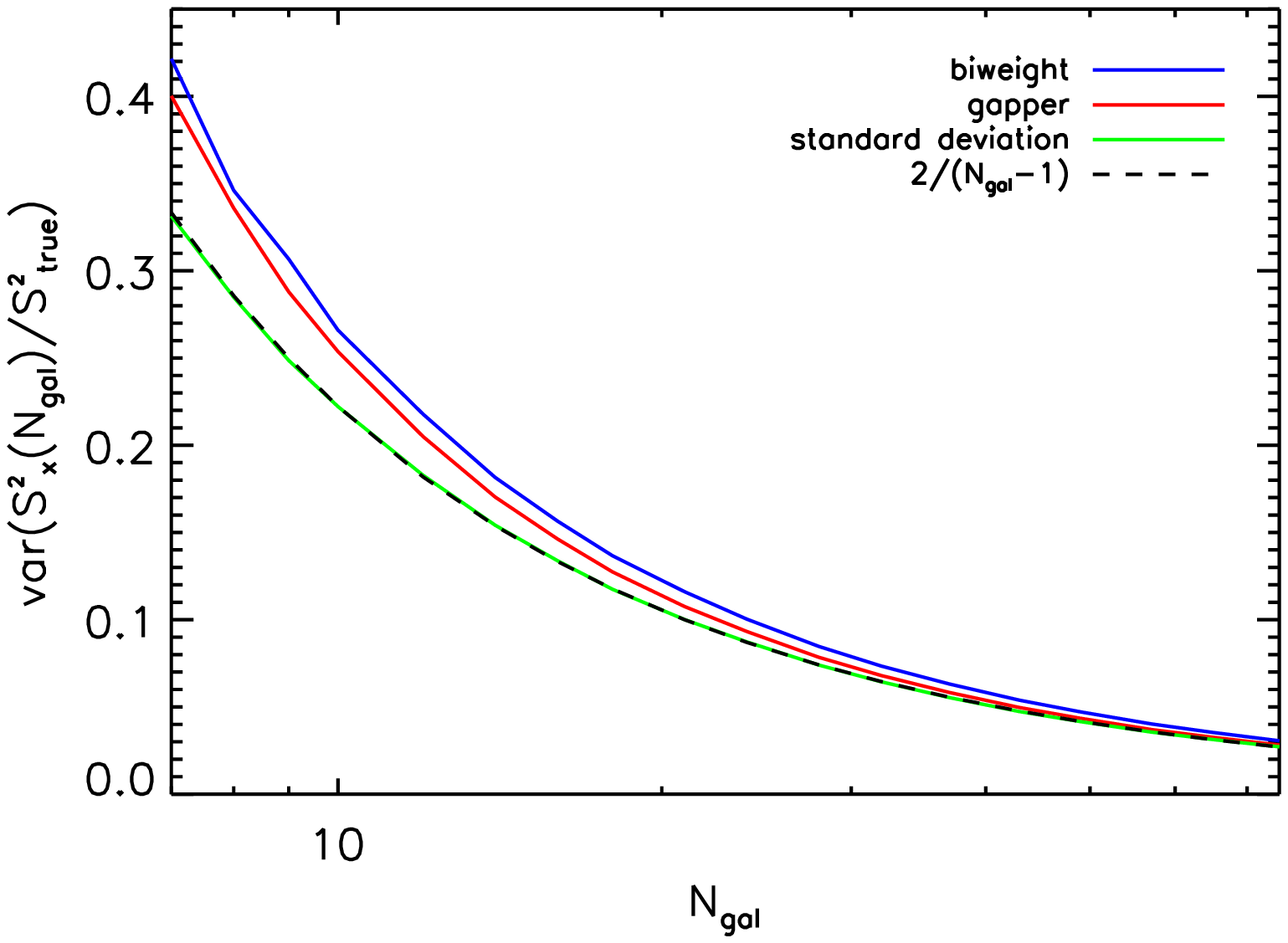}
\end{center}
\caption{Mean (left) and variance (right) of $\Sx^2(\ngal)$ as a
  function of the number of galaxies $\ngal$, for the case of Gaussian
  random velocities. Green, red, and blue lines represent respectively the behaviour of the standard
  deviation, gapper, and biweight estimators. }
  \label{fig:gauss_2}
\end{figure*}

In this case of Gaussian velocity distributions, it is also interesting to test the
response of the square of these three estimators $\left<\Sx^2(\ngal)/S_{\rm
  true}^2\right>$, which in the case of the standard deviation corresponds to
the variance. From a theoretical point of view, the variance is one of
the ordinary moments of a distribution, and is not
biased. Fig.~\ref{fig:gauss_2} presents the results, confirming this
well-known behaviour for the standard deviation, and showing that the
other two estimators are clearly biased in this low $\ngal$ regime. Also, the variance of
the variance follows the expected analytic dependence for the standard
deviation,
\begin{equation}
var(\Sx^2(\ngal))=\frac{2 S_{\rm true}^4}{(N_{\rm gal}-1)}.
\label{eq:v_sig_teoric}
\end{equation}
We note that both the gapper and biweight have a similar dependence of
the variance on $\ngal$, being the biweight the estimator presenting a
higher variance.

Following the methodology applied in Sect.~\ref{sec:bias}, we can
construct the unbiased scale estimators $\Sxprime$. We have checked
that this new set of corrected estimators $\Sxprime$, defined as in
equation~\ref{eq:un_S} and using the parameters listed in
table~\ref{table:param_un}, provide unbiased estimates for the
velocity dispersion.

\subsection{Mass estimators}

% MASS
%
\begin{figure*}
\begin{center}
\includegraphics[width=0.49\textwidth]{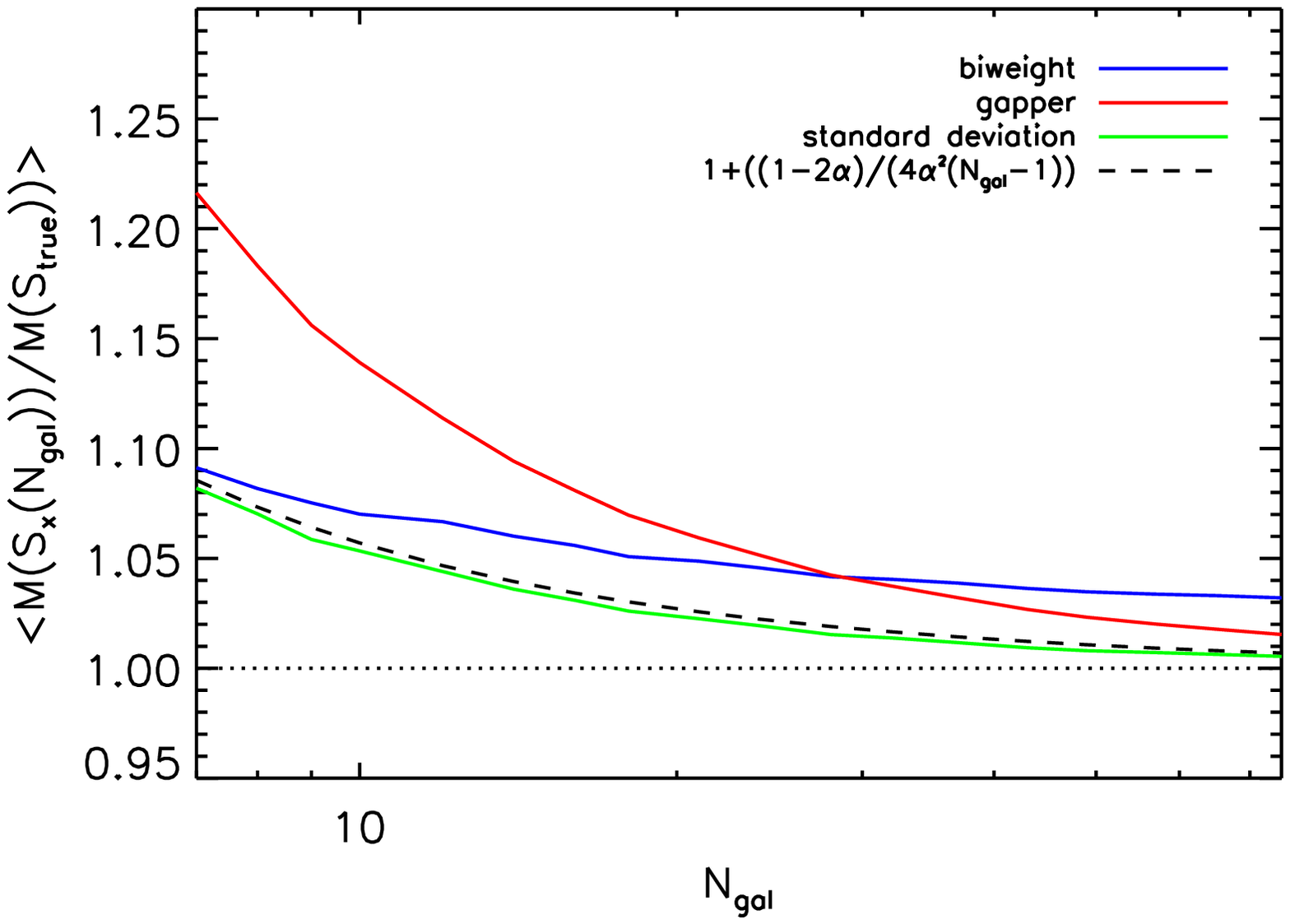}
\includegraphics[width=0.49\textwidth]{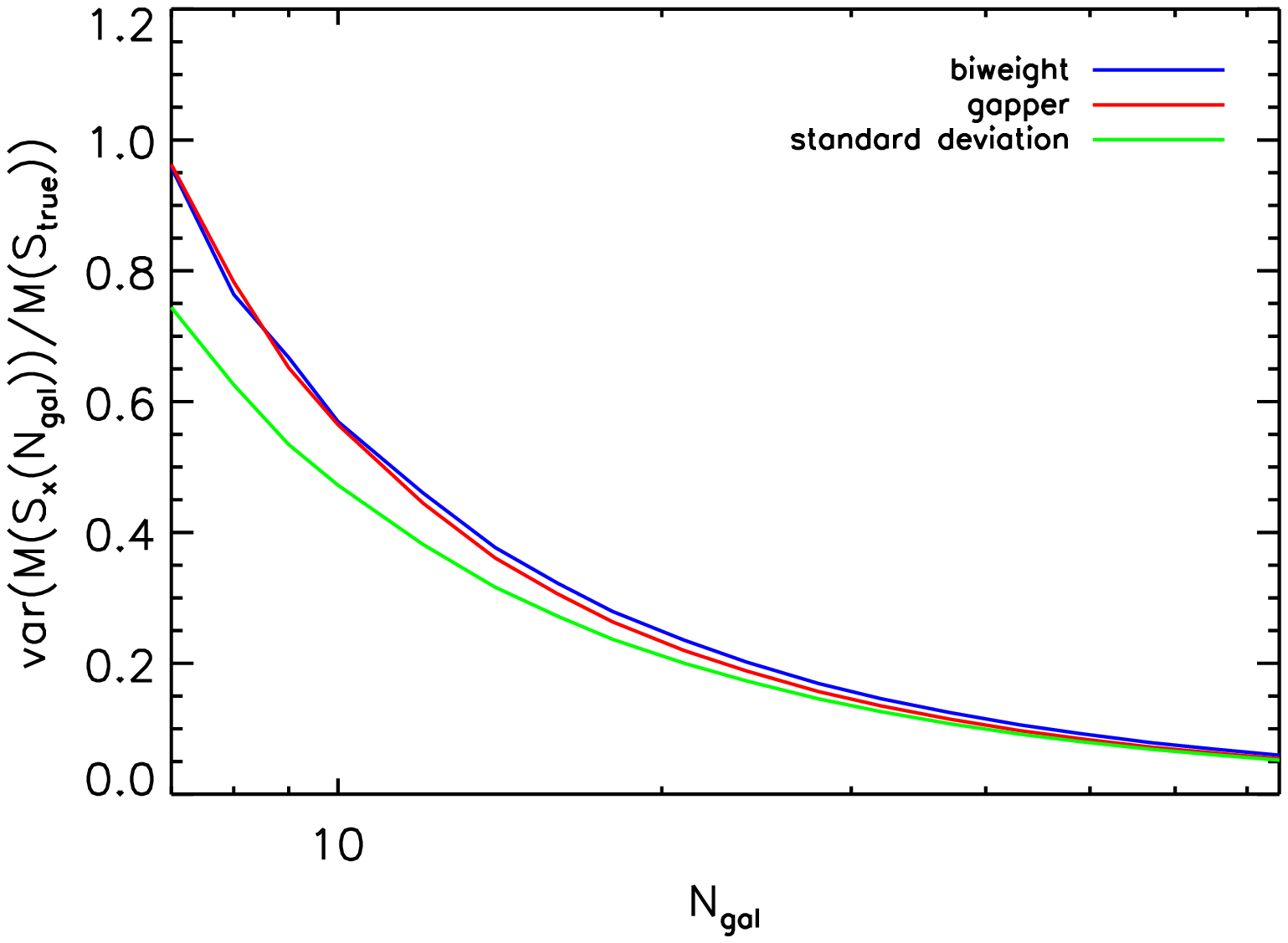}
\end{center}
\caption{Mean (left) and variance (right) of $M\left(\Sx(\ngal)
  \right)/M(S_{\rm true})$ as a function of $\ngal$, for the case of Gaussian random velocities. The dashed line in the left panel
  represents the expected analytic formula for the dependence of $F(S_{\rm std})$ on
  $\ngal$ for the standard deviation described in Appendix~\ref{apendix:ap1}. Green, red, and blue lines represent respectively the behaviour of the standard
  deviation, gapper, and biweight estimators.  }
  \label{fig:m_b_gauss_f}
\end{figure*}

We now discuss the behaviour of the mass estimators in the case of
Gaussian velocity distributions. Here, the reference
true mass, $M(\Strue)$, is calculated applying eq.~\ref{eq:smg}, using
the parameters $A=1177.0$\,km\,s$^{-1}$ and $\alpha=0.364$ from \cite{munari13}.
Fig.~\ref{fig:m_b_gauss_f} illustrates the bias for the
  uncorrected mass estimator $M\left(\Sstd(\ngal) \right)/M(\Strue)$. In
general, the dependence on $\ngal$ for all mass estimators is very similar to that
in Fig.~\ref{fig:m_b_sim_f}. In particular, the standard deviation mass estimator follows the theoretical analytic
form presented in eq.~\ref{eq:ap5}.  
The right panel in the same figure shows the variance of
$M\left(\Sx(\ngal) \right)/M(\Strue)$. Although we note a slightly
higher variance at low $\ngal$, the shape of variance profiles are
similar to those presented in figure~\ref{fig:m_b_sim_f} for the real
cluster simulations.

\begin{table}
\caption{Best-fit parameters of the functions $\left<M\left( \Sx(\ngal) \right)/M(S_{\rm
    true})\right>$ for Gaussian velocity distributions. }
\label{table:Fs_par} 
\centering 
\begin{tabular}{c c c c} 
\hline\hline 
\noalign{\smallskip}
\, & $BWT$ & $GAP$ & $STD$ \\
\hline
\noalign{\smallskip}
$E$ & $2.22\pm0.02$ & $1.505\pm0.007$ & $2$\\ 
$F$ & $0.023\pm0.001$ & $0.0013\pm0.0005$ & $0$\\
$\gamma$ & $0.82\pm0.04$ & $ 1.088\pm0.009 $ & $ 1 $ \\
\noalign{\smallskip}
\hline 
\end{tabular}
\end{table}
\begin{table}
\caption{Best-fit parameters of the functions $\left<M\left(\Sxprime(\ngal)
  \right)/M(\Strue)\right>$ for Gaussian velocity distributions. }
\label{table:Fs'_par}
\centering
\begin{tabular}{c c c c} 
\hline\hline 
\noalign{\smallskip}
\, & $BWT$ & $GAP$ & $STD$ \\
\hline
\noalign{\smallskip}
$E'$ & $1.28\pm0.01$         & $1.504\pm0.007$     & $1.521\pm0.007$\\ 
$F'$ & $0.0078\pm0.0008$ & $0.0032\pm0.0005$ & $0.0009\pm0.0005$\\
$\gamma^\prime$ & $1.23\pm0.01$         & $ 1.088\pm0.009 $   & $ 1.083\pm0.008 $ \\
\noalign{\smallskip}
\hline 
\end{tabular}
\end{table}

Bias-corrected mass estimators for Gaussian distributions, based both on the plain
  estimators $\Sx$, or their unbiased counterparts $\Sxprime$, can be obtained using
equation~\ref{eq:M_M'_eq} with the parameters listed in Tables~\ref{table:Fs_par} and \ref{table:Fs'_par}, respectively.  
Similarly to the case of true velocity distributions, in this case of
Gaussian velocities the biweight has also the largest variance,
whereas standard deviation has the lowest one, with
almost the same behaviour as the gapper.

\section{Analytical fit to the variance of the different estimators }
\label{apendix:ap4}

Sect.~\ref{subsec:def} and equations~\ref{eq:un_S}, \ref{eq:M_M'_eq}
and \ref{eq:M_M"_eq} present the definitions of different estimators
for the velocity dispersion and the mass for galaxy clusters.  
For all these cases, figures~\ref{fig:n_sigma}, \ref{fig:n_sigma2},
\ref{fig:m_b_sim_f} and \ref{fig:m_b_sim_M} showed the mean and
variance of the three estimators, as a function of $\ngal$. Analytical
expressions for the corrected estimators were given in the main text. 

In some applications, it is useful to have also an analytical fit to
the variance of those estimators. Here we provide this fit, using a
common expression for all cases with only two free parameters:
\begin{equation}
Var = \frac{\epsilon}{4(\ngal -1)^\beta}. 
\end{equation}
Table~\ref{table:fit_var} contains the the best-fit values for the $\epsilon$ and
$\beta$ parameters for all those estimators. In general, the standard
deviation estimator has the lowest variance in all cases, being the
variance of the gapper also very close but slightly larger. The
biweight estimator has a stronger dependence on $\ngal$, specially in
the low-$\ngal$ regime.

\begin{table*}
\caption{Coefficient of the numerical fit to the dependence of the
  variance as a function of $\ngal$ for different estimators.  }
\label{table:fit_var}
\centering
\begin{tabular}{c c c c} 
\hline\hline 
\noalign{\smallskip}
Estimator & BWT & GAP & STD \\
\hline 
\noalign{\smallskip}
$Var(\Sx )$& $\epsilon = 3.69 \pm 0.10$ &  $\epsilon = 2.42 \pm 0.05$ &  $\epsilon = 2.05 \pm 0.06$ \\
                  & $\beta = 1.24 \pm 0.01$  & $\beta = 1.15 \pm 0.01$  & $\beta = 1.11 \pm 0.01$  \\
\hline 
\noalign{\smallskip}
$Var(\Sxprime )$& $\epsilon = 4.63 \pm 0.16$ &  $\epsilon = 2.42 \pm 0.05$ &  $\epsilon = 2.36 \pm 0.06$ \\
                           & $\beta = 1.33 \pm 0.02$  & $\beta = 1.15 \pm 0.01$  & $\beta = 1.15 \pm 0.01$  \\
\hline 
\hline
\noalign{\smallskip}
$Var(M(\Sx) )$& $\epsilon = 31.7 \pm 1.1$ & $\epsilon=  34.4 \pm 1.0$& $\epsilon=  21.2 \pm 0.5$ \\
                      & $\beta =  1.26 \pm 0.02$ & $\beta=  1.32 \pm 0.013$&$\beta=  1.20\pm 0.01$\\
\hline
\noalign{\smallskip}
$Var(M(\Sxprime) )$& $\epsilon=  61.0 \pm 3.7$& $\epsilon= 32.9 \pm 0.9$&$\epsilon=  32.1 \pm 0.8$  \\
                               & $\beta= 1.50 \pm 0.03$& $\beta = 1.33 \pm 0.01$&$\beta=  1.33 \pm 0.01$\\
\hline
\noalign{\smallskip}
$Var(M'(\Sx) )$& $\epsilon= 23.7 \pm 0.8$& $\epsilon= 16.7 \pm 0.5$&$\epsilon=16.3 \pm 0.5  $  \\
                               & $\beta= 1.21 \pm 0.01$& $\beta =1.13 \pm 0.01 $&$\beta=  1.13 \pm 0.01 $\\
\hline
\noalign{\smallskip}
$Var(M'(\Sxprime) )$& $\epsilon= 23.9 \pm 0.8 $& $\epsilon= 16.9 \pm 0.5$&$\epsilon=16.2 \pm 0.5   $  \\
                               & $\beta= 1.21 \pm 0.02$& $\beta = 1.13\pm 0.01 $&$\beta=1.13 \pm 0.01  $\\
\noalign{\smallskip}
\hline\hline
\end{tabular}
\end{table*}

\end{appendix}

\end{document}